\documentclass{article}

\usepackage{arxiv}

\usepackage[utf8]{inputenc} 
\usepackage[T1]{fontenc}    
\usepackage{hyperref}       
\usepackage{url}            
\usepackage{booktabs}       
\usepackage{amsfonts}       
\usepackage{nicefrac}       
\usepackage{microtype}      
\usepackage{lipsum}
\usepackage{graphicx}
\graphicspath{ {./images/} }

\usepackage{soul}
\usepackage[super]{natbib}
 \bibpunct[, ]{(}{)}{,}{a}{}{,}%
\usepackage[flushleft]{threeparttable} 
\usepackage{subcaption}
\usepackage{mathtools}
\usepackage{algorithm}
\usepackage{algpseudocode}
\usepackage{enumitem}

\setlist[enumerate]{nosep, leftmargin=1.4cm}
\newlist{steps}{enumerate}{1}
\setlist[steps, 1]{nosep, leftmargin=2cm, label = \textbf{Step\arabic*}}
\usepackage{xcolor}
\usepackage{physics} 
\usepackage{booktabs}
\usepackage{multirow}

\usepackage{hyperref}

\usepackage{setspace}
\makeatletter
\renewcommand\subsubsection{\@startsection{subsubsection}{3}{\z@}%
{-3.25ex\@plus -1ex \@minus -.2ex}%
{1.5ex \@plus .2ex}%
{\normalfont\normalsize\bfseries}}
\makeatother

\usepackage{amsthm,amsfonts,amssymb,amsmath,graphicx}
\usepackage{soul}

\newtheorem{proposition}{Proposition}

\newtheorem{definition}{Definition}

\title{Evidence and quantification of cooperation of driving agents in mixed traffic flow}

\author{
Di Chen \\
  Department of Civil and Environmental Engineering\\
  University of California, Davis\\
  Davis, CA 95616 \\
  \texttt{diichen@ucdavis.edu} \\
   \And
 Jia Li \\
  Department of Civil and Environmental Engineering\\
  Washington State University\\
  Pullman, WA 99163 \\
  \texttt{jia.li1@wsu.edu} \\
  \And
 H. Michael Zhang* \\
  Department of Civil and Environmental Engineering\\
  University of California, Davis\\
  Davis, CA 95616 \\
  \texttt{hmzhang@ucdavis.edu} \\
  * Corresponding author \\
}

\begin{document}
\maketitle
\begin{abstract}
Cooperation is a ubiquitous phenomenon in many natural, social, and engineered systems with multiple agents. Understanding the formation of cooperation in mixed traffic is of theoretical interest in its own right, and could also benefit the design and operations of future automated and mixed-autonomy transportation systems. However, how cooperativeness of driving agents can be defined and identified from empirical data seems ambiguous and this hinders further empirical characterizations of the phenomenon and revealing its behavior mechanisms. Towards mitigating this gap, in this paper, we propose a unified conceptual framework to identify collective cooperativeness of driving agents. This framework expands the concept of collective rationality from our recent model~\citep{li2022equilibrium}, making it empirically identifiable and behaviorally interpretable in realistic (microscopic and dynamic) settings. This framework integrates mixed traffic observations at both microscopic and macroscopic scales to estimate critical behavioral parameters that describe the collective cooperativeness of driving agents. Applying this framework to NGSIM I-80 trajectory data, we empirically confirm the existence of collective cooperation and quantify the condition and likelihood of its emergence. This study provides the first empirical understanding of collective cooperativeness in human-driven mixed traffic and points to new possibilities to manage mixed autonomy traffic systems.

\end{abstract}


\section{Introduction}

Cooperation is a ubiquitous phenomenon in many natural, social, and engineered systems that involve interactions of multiple agents \citep{sachs2004evolution}. Animals, such as birds, fishes, and ants, can organize themselves into groups and behave cooperatively as a whole -- birds fly in flocks to save energy, fishes swim in schools to resist attacks, and ants form lanes to minimize congestion \citep{parrish2002self,couzin2003self,couzin2003self2}. In these examples, cooperation emerges in the form of spatial patterns and structures. Researchers have devoted much efforts towards understanding the behavioral mechanisms of such cooperation (see e.g. \cite{bernasconi1999cooperation,mccreery2014cooperative}).

Characterizing and understanding cooperation in mixed traffic is of interest and significance for two reasons. From a theoretical standpoint, a central problem in traffic flow theory is to understand how traffic agents interact with each other in microscopic scale and the emergence of collective behaviors in macroscopic scale in traffic flow. While the micro-macro linkage in traffic flow has been extensively studied, most of them focused on the longitudinal behaviors of identical driving agents \citep{lee2001macroscopic,jin2016equivalence, chiarello2021statistical}. Gaining insights into how cooperation emerges in mixed traffic flow will expand and deepen our understanding of both longitudinal and lateral behaviors of mixed traffic. Pragmatically, in mixed autonomy transportation systems, important intelligent transportation applications such as cooperative adaptive cruise control (CACC) are often built on the idea of coordination between driving agents (i.e. automated vehicles) or driving agents with infrastructures (e.g. traffic signals), where it is usually assumed the agents share a common objective through system performance functions \citep{liu2019traffic,li2022cooperative}. It is natural to ask when driving agents are self-interested, no longer share a common objective, and accordingly lacks external coordination, whether ``cooperation'' can spontaneously emerge from their interactions. To answer this question, in the first place, there needs to be a measurable and interpretable description of cooperation in mixed traffic. 

Nonetheless, characterizing cooperation accurately in mixed traffic is not a trivial task, due to the lack of a quantifiable and behaviorally sound definition of cooperativeness in mixed traffic, which could behave subtlely. Roughly, by ``\ul{cooperation}'', we refer to the formation of spatiotemporal structures in mixed traffic that benefit overall traffic flow, and by ``\ul{cooperativeness}'', we refer to the capability of driving agents to form cooperation. When, the cooperativeness of a class of agents, instead of an individual one, is concerned, we call it ``\ul{collective cooperativeness}''. Specifically, the gap exists due to several reasons. First, though empirical characteristics, such as spontaneous platooning (clustering) \citep{neubert1999single,helbing2001traffic,al2009platooning,yang2015empirical}, lane choice and lane changing behavior \citep{amin2005variation,duret2012lane,reina2020lane,espadaler2023empirical}, herding behaviors \citep{kamrani2020applying,nagahama2021detection}, are indicative of whether or not a driving agent is cooperative, they each only provide a partial characterization of behavior and choice of driving agents, rather than a holistic one. Second, these characteristics are descriptive in nature and miss capturing the interplay between driving agents explicitly. Given that cooperativeness intrinsically involves more than one agents, such characteristics are insufficient to describe or interpret the formation of cooperation. Third, real-world data are usually noisy, and behaviors of mixed traffic are determined not only by cooperativeness of agents, but also car-following dynamics, lane policy, and road geometry. Recipes need to be carefully designed to empirically delineate the cooperativeness of agents from other confounding factors.

This paper attempts to bridge this gap. We will focus on equilibrium relation of mixed traffic and formulate a computable definition of ``collective cooperativeness'' for driving agents in mixed traffic and develop a new framework to identify it from vehicle trajectory data. The core idea stems from our recent work ~\citep{li2022equilibrium}, which proposed an analytical two-player bargaining game model to capture the agent interactions in mixed autonomy traffic. However, while this model offers insights into the possible agent interactions and the resulting equilibria, it makes substantial simplified assumptions that need to be relaxed in a realistic empirical setting: \cite{li2022equilibrium} treats mixed traffic as continuum at equilibrium and focuses on the lateral interactions of two classes of agents, without explicitly considering longitudinal driving dynamics, lane changes, and lane setting. Consequently, the bottom-up connections between this model and microscopic measurements are not clear and it is necessary to formulate a new definition of ``cooperativeness'' to make empirical identification possible. 

Specifically, the major contributions of this paper include:
\begin{itemize}
    \item[] \#1 (Conceptual): We introduce three representation spaces and adapt the theoretical concept of collective cooperativeness to make it empirically quantifiable and behaviorally interpretable in real-world scenarios.
    \item[] \#2 (Technical): We develop a framework to empirically quantify collective cooperativeness of mixed traffic, leveraging multi-scale information of vehicle trajectory data.
    \item[] \#3 (Empirical): We conduct a case study using real-world trajectory data of mixed human-driven traffic and show the existence of collective cooperation and when it is attained.
\end{itemize}

The rest of the paper is organized as follows. In Section~\ref{sec:lit}, we delve into equilibrium models of mixed traffic flow and discuss how collective cooperativeness is implicitly or explicitly embodied in such models. In Section~\ref{sec:prelim}, we present major concepts and theoretical findings from our recent work~\citep{li2022equilibrium}, which serve as a foundation for this study. In Section~\ref{section:method}, we present the extension of collective cooperativeness concept to physical space and propose a framework that identifies collective cooperation of traffic agents in realistic multilane freeway settings. In Section~\ref{section:empirical}, we present a case study and demonstrate application of the framework using real-world NGSIM I-80 trajectory data. We conclude the paper with an outlook of future research in Section~\ref{section:conclusion}.

\section{Equilibrium traffic models and collective cooperativeness}\label{sec:lit}

Early models of mixed traffic are multi-class extensions to the classical LWR model~\citep{lighthill1955kinematic, richards1956shock} based on macroscopic (fluid level) descriptions. These models \citep{wong2002multi,logghe2003dynamic,jin2004multicommodity,ngoduy2007multiclass,van2008fastlane,levin2016multiclass,shi2021constructing} usually do not directly capture agent-level behaviors and their interactions, so we call them ``descriptive mixed traffic models''. Central to these models are the speed functions (or equivalently, fluxes) of different class of agents. The general form reads,
\begin{equation}\label{eq_des}
u_i=U_i(\rho_1,\rho_2), \ i=1,2
\end{equation}
where $u_i$ and $U_i(\cdot,\cdot)$ are respectively speeds and speed functions of class-$i$ agents. To determine $U_i(\cdot,\cdot)$, these models usually adopt the familiar ``free-flow'' and ``congestion'' notions from single-class traffic, and specify free-flow and congestion conditions based on the values of class density $\rho_1$ and $\rho_2$. In the free flow regime, the two classes of agents are independent and attain their respective free flow speeds, while in the congestion regime, speeds of the agents are synchronized.

The major limitation of the above approach lies in that the free flow and congestion regimes are exogenously specified, which to a certain extent, is arbitrary. The root cause is that agent interactions are not explicitly considered in those models, and thus they lack a microscopic behavioral foundation. Another class of models (which we call ``behavioral mixed traffic models'') remedy this problem, by modeling interactions of agents endogenously. One early effort is by \citet{daganzo1997continuum}, who modeled a two-lane freeway with one special lane. The traffic consists of two classes of vehicles that have identical fundamental diagrams, but only one class can access both lanes. In this case, that class of agents makes the decision of whether or not to use the general purpose lane. The notion of user equilibrium is used to describe the choice of this class.  \citet{daganzo2002behavioral_a,daganzo2002behavioral_b} further studied more complicated lateral interactions between two classes of traffic agents, called ``rabbits'' and ``slugs'', in a general multi-lane freeway setting. \citet{daganzo2002behavioral_a} proposed a macroscopic behavioral theory that explains some complex phenomena in traffic flow. In this theory, all agents are self-interested (also known as ``self-serving"), though they have different lane preferences. The notion of lane is later generalized to a division factor, a description of how lateral road spaces are allocated among different classes, and considered in subsequent mixed traffic flow models. In this spirit, \citet{logghe2008multi} developed an equilibrium model for multi-class traffic based on the user equilibrium principle. A division factor was introduced to model the interactions between two classes of agents. Building upon a similar idea, \citet{qian2017modeling} proposed another mixed traffic model and offered a new interpretation of the division factor as the proportion of road spaces that different classes of traffic streams compete for. The key difference of behavioral mixed traffic models from the previous ones lies in the introduction of a lateral dimension, which describes total lateral space, or ``road share''. Interactions of agents are captured by considering the split of road share between them. This way, the congestion regime is established in a more behaviorally sound way. To see the difference of these models from the previous ones, we note that \cite{logghe2008multi} and \cite{qian2017modeling} share the same general form,
\begin{equation}\label{eq_beh}
    u_i=V_i(\rho_i/p_i), \ i=1,2
\end{equation}
where $V_i(\cdot)$ is the nominal speed function and $p_i$ is the road share assigned to class-$i$, which is a function of $\rho_1$ and $\rho_2$, i.e. $p_i=p_i(\rho_1,\rho_2)$. Compared to (\ref{eq_des}), the main advantage of (\ref{eq_beh}) lies in its clearer behavior interpretation and less ambiguity in specifying the free-flow and congestion regimes. Interactions of agents are fully encapsulated in the definition of road share functions.

The cooperativeness concept naturally emerges in the second class of models, because split of road share is not necessarily unique, but rather, depends on how agents interact. \citet{logghe2008multi} touched upon this point. It is postulated that trucks tend to organize themselves into as few lanes as possible as long as their speed is not compromised, while cars take the remaining lanes. In this case, trucks are cooperative. This assumption, however, is strong. For instance, trucks and cars may compete for road share and cannot reach a consensus on how road share should be splitted between them. Out of this observation, \cite{li2022equilibrium} propose a generalized model of mixed autonomy traffic to capture the ``bargaining'' of two classes of agents for road share and put forward the concept of ``collective rationality/cooperation'', Pareto-efficient equilibria. The model showed that collective cooperation (rigorous definition to be given in Section~\ref{sec:prelim}) can emerge endogenously in mixed autonomy traffic.

\begin{table}[h]
    \centering
    \caption{Comparison of representative multi-class traffic flow models}
    \label{tab:studies_comparison}
    \begin{tabular}{p{5.6cm}ccccc}
        \toprule
        Study & Longitudinal & Lateral$^1$ & Behavioral & Cooperative$^2$ & Empirical \\
        \midrule
        \citet{daganzo1997continuum} & $\bullet$ & $\bullet$ &  $\bullet$ & - & - \\
			\citet{daganzo2002behavioral_a,daganzo2002behavioral_b} & $\bullet$ & $\bullet$ & $\bullet$ & - &- \\
			\citet{wong2002multi} & $\bullet$ & - & - & - & - \\
			\citet{logghe2003dynamic} & $\bullet$ & - & - & - & $\bullet$ \\
			\citet{jin2004multicommodity} & $\bullet$ & - & - & - & - \\
			\citet{ngoduy2007multiclass} & $\bullet$ &  - & - & - & -\\
			\citet{van2008fastlane} & $\bullet$ &  - & - & - &  $\bullet$ \\
			\citet{logghe2008multi}  & $\bullet$ & $\bullet$ & $\bullet$ & $\circ$ & - \\
			\citet{levin2016multiclass} & $\bullet$ &  - & - & - & - \\
			\citet{chen2017towards} & $\bullet$ & $\circ$ & - & - & - \\
			\citet{ghiasi2017mixed} & $\bullet$ & $\circ$ & - & - & - \\
			\citet{qian2017modeling} & $\bullet$ & $\bullet$ & $\bullet$ & $\circ$ & $\bullet$ \\
			\citet{huang2019game} & $\bullet$ &  - & - & - & - \\
			\citet{qin2021lighthill} & $\bullet$ & - & - & - & - \\
                \citet{shi2021constructing} & $\bullet$ & - & - & - & $\bullet$ \\
			\citet{li2022equilibrium} & $\bullet$ & $\bullet$ & $\bullet$ & $\bullet$ & -\\
        \bottomrule
    \end{tabular}
    \begin{tablenotes}
        \small
        \item[1] 1. ``$\bullet$'' means lane choice decisions are endogenous; ``$\circ$'' means exogenous lane policies are imposed.
        \item[2] 2. ``$\bullet$'' means cooperation endogenously arises; ``$\circ$'' means cooperation is exogenously assumed.
    \end{tablenotes}
\end{table}

To further clarify the research gap, we summarize and compare major mixed traffic flow models in Table~\ref{tab:studies_comparison}. Five dimensions of these models are compared. The ``longitudinal'' column means whether the longitudinal behaviors (i.e., car-following) of agents are modeled. The ``lateral'' column means whether the distribution of agent classes across the lanes is modeled. Two types of lateral models are considered. In the first type of model, the lateral distribution of traffic agents is determined by exogenous lane policies. In the second type of model, the agents' lane choices are captured as endogenous decisions. The ``behavioral'' column describes whether traffic agents make lane choice decisions on their own, rather than following a given lane policy. Note that because of the subtle difference, a model with the lateral dimension does not necessarily consider the lateral behaviors of agents. Two examples are \citet{chen2017towards} and \citet{ghiasi2017mixed}. The ``cooperative'' column describes whether a model endogenously models the cooperation of traffic agents. Last, ``empirical'' means whether a model is validated against real-world data. This table shows that while the concept of cooperation not yet fully studied in mixed traffic flow, it ties closely to the lateral interactions of driving agents and finds its root in behavioral mixed traffic models. Nonetheless, while recent models \citep{logghe2008multi,qian2017modeling,li2022equilibrium} have paved the way, a practical definition and empirical identification of cooperativeness for driving agents are yet to be developed.

\section{Prior definition and properties of collective cooperativeness}\label{sec:prelim}


In this section, we briefly recapitulate the original concept of collective cooperativeness in mixed traffic, proposed in our previous work~\citep{li2022equilibrium}. The concept is based on a two-player bargaining game model of mixed autonomy traffic. The model treats each class of traffic agents as a player and their interactions are encapsulated as a collective bargaining process: the agents negotiate to determine the allocation of lateral road share and settle to certain Nash equilibria (NE). One noteworthy property of the model is that it contains multiple NE, a subset of which is Pareto-efficient. Collective cooperativeness arise when the Pareto-efficient NE are attained, which serve as the conceptual foundation of this paper. 

The major notations to be used are summarized in Table~\ref{tab_notation}.

\begin{table}[h]
    \centering
    \caption{Major notations in the paper}
    \label{tab_notation}
    \begin{tabular}{ll}
        \toprule
        Notation & Explanation \\ 
        \midrule
        $i,j$ & class index of traffic agents\\
    $a_j,b_j$ & scaling parameters\\
    $\rho_i$ & macroscopic traffic density\\
    $\rho_{ij}^*$ & microscopic traffic density \\ 
    $u_i(\cdot)$ & nominal speed function \\
    $u^*(\cdot,\cdot)$ & 1-pipe speed \\
    $p_i$ & road share bid  \\
    $p_i^*$ & minimum road share \\
        \bottomrule
    \end{tabular}
\end{table}

To be precise, in the two-player bargaining game model of~\citet{li2022equilibrium}, self-interested agents aim to maximize their travel speed by negotiating the road share (i.e., the proportion
of lateral space) they take up. The pay-off function to the players is written as,
\begin{equation}\label{eq:payoff}
U_i(\rho_1,\rho_2,p_1,p_2) = 
    \begin{cases}
    \begin{aligned}
     &u^*(\rho_1, \rho_2) & \text{if $p_1+p_2 > 1$}, \\
    &u_i(\rho_i/p_i) & \text{if $p_1+p_2 \leq 1$}
    \end{aligned}
    \end{cases}
    i=1,2
\end{equation}
Here, $u^*(\cdot,\cdot)$ denotes the one-pipe speed, meaning that both classes move at the same speed. The term $p_i \in [0,1]$ represents the proportion of lateral road space that class $i$ agents bid for. The total road share bid, $p_1+p_2$, determines the traffic regimes: when both classes agree on their respective road share bids, they travel in separate lanes, leading to a 2-pipe regime ($p_1+p_2 \leq 1$); conversely, if an agreement is not reached, the two classes will mix together, resulting in 1-pipe regime ($p_1+p_2 > 1$).

To derive Nash equilibria of the bargaining game, the concept of minimum road share was proposed,
\begin{equation}\label{eq:p1*_p2*}
    p_1^* = \frac{\rho_1}{u_1^{-1}(u^*)}, \ p_2^* = \frac{\rho_2}{u_2^{-1}(u^*)}
\end{equation}
where $p_i^*$ is the minimum road share taken by class $i$. The total minimum road share, $p_1^*+p_2^*$, determines the type of NE (i.e., Pareto-efficient or non-Pareto-efficient) attained in the bargaining game.

To be more specific, $p_1^* + p_2^* > 1$ means two classes of agents are unable to agree on a split of road share, and in this case, fully mix (1-pipe) state is the only NE, and this NE is Pareto-efficient. Conversely, when $p_1^* + p_2^* \leq 1$, the traffic has two types of NEs, one being fully mix, and the other being fully separate (2-pipe). In this case, the 2-pipe NEs are Pareto-efficient. Collective cooperativeness is attained when the system reaches 2-pipe NEs, characterized by $p_1^* + p_2^* \leq 1$.

In the following, we will review major conclusions on the Pareto efficient equilibria that theoretically defines collective cooperativeness, and refer readers to~\citet{li2022equilibrium} for complete definitions and proofs.

\textbf{1-pipe equilibrium}

When mixed traffic system settles to a 1-pipe equilibrium, all agents move at the same speed, denoted as the 1-pipe speed $u^*$. The governing equation for $u^*$ follows,

\begin{equation}\label{eq_newgov}
\frac{1}{\rho_{tot}}\sum_{i=1}^2\sum_{j=1}^2\frac{\rho_{i}\rho_j }{u_{ij}^{-1}(u^*)}=\frac{1}{\rho_{tot}}\sum_{i=1}^2 \rho_i \left(\sum_{j=1}^2 \frac{\rho_j }{u_{ij}^{-1}(u^*)} \right)=1
\end{equation}
where $\rho_i$ is the density for class $i$ agents, $\rho_{tot}=\rho_1+\rho_2$ is the total system density. The nominal speed function $u_{ij}(\rho)$ represents the speed of a class $i$ agent when it follows a class $j$ agent, and $u_{ij}^{-1}(u^*)$ is its inverse form. 

For analytical tractability, \citet{li2022equilibrium} considered a special case of the equilibrium speed $u^*$, assuming that all speed functions follow the form: $u_{ij}(\rho)=u(\rho/a_{ij})$. In this expression, $a_{ij}$ is a scaling parameter capturing the dependency of vehicle speed on vehicle classes, reflecting type sensitivity between agents, and $u(\cdot)$ denotes a reference speed-density relation. This equations assumes that both classes are endowed with a same nominal speed function $u(\cdot)$. In this empirical study, considering the heterogeneity between vehicle classes in real-world, we refine this definition to incorporate two nominal speed functions,
\begin{equation}\label{eq:scaling}
u_{1j}(\rho) = u_1(\rho/a_j), u_{2j}(\rho) = u_2(\rho/b_j), j=1,2
\end{equation}
where $u_1(\cdot)$ and $u_2(\cdot)$ are the nominal speed-density function for class 1 and class 2 vehicles, respectively, and $a_j, b_j$ are the scaling parameters for each class.
This simplification leads to $u_{1j}^{-1}(u^*)=a_{j} u_1^{-1}(u^*)$ and $u_{2j}^{-1}(u^*)=b_{j} u_2^{-1}(u^*)$, and the 1-pipe speed in (\ref{eq_newgov}) is therefore reduced to,
\begin{equation}\label{eq:1pipe_speed}
\frac{1}{\rho_{tot}} 
\left(\frac{\rho_1}{u_1^{-1}(u^*)} \sum_{j=1}^{2} \frac{\rho_j}{a_j} + \frac{\rho_2}{u_2^{-1}(u^*)} \sum_{j=1}^{2} \frac{\rho_j}{b_j}\right) = 1
\end{equation}
This equation provides an implicit function of $u^*(\rho_1,\rho_2)$, which can be solved to obtain the 1-pipe equilibrium speed (the original form can be found in Eq.(11) in~\citet{li2022equilibrium}). To solve this implicit function, we require knowledge of the nominal speed functions $u_j(\cdot)$, macroscopic densities $\rho_j$, and scaling parameters $a_j$ and $b_j$. The estimation of these functions and parameters is part of the cooperativeness identification framework of this study, which will be elaborated in the following sections.

\textbf{2-pipe equilibria}

At 2-pipe Pareto efficient equilibria, we have $p_1^*+p_2^* \leq 1$. This leads to a leftover portion of road share arising from the interactions between agents. We refer to such extra road share as the \textbf{cooperation surplus}, denoted as $s$. This surplus represents an additional road share agents can utilize to be better off than what is achievable in the 1-pipe equilibrium. The formal definition can be seen in Definition~\ref{def:surplus}. A positive cooperation surplus is a necessary condition of attaining collective cooperativeness. 

\begin{definition}[Cooperation surplus]\label{def:surplus}
We call the road share left from players' cooperation as cooperation surplus, $s \coloneqq 1-p_1^*-p_2^*$.
\end{definition}

Moreover, while the one-pipe equilibrium has proven to be unique, this is not the case for 2-pipe equilibria. \citet{li2022equilibrium} further considered a scenario that leads to a unique 2-pipe equilibrium, which reflects how the benefit of cooperation is allocated between two classes of agents. In this equilibrium, the distribution of road share between each vehicle class becomes, 
\begin{equation}
\begin{cases}
\begin{aligned}
    &p_1 = p_1^* +\lambda(\rho_1, \rho_2) s \\
    &p_2 = p_2^* +(1-\lambda(\rho_1, \rho_2)) s
\end{aligned}
\end{cases}
\end{equation}
where $\lambda$ is denoted as the  \textbf{surplus split factor}, representing the way of splitting the cooperation surplus. It is formally defined as,
\begin{definition}[Surplus split factor]\label{def:surplus_split}
The effective split of the cooperation surplus by the two players is called the surplus split factor, denoted as $\lambda(\rho_1, \rho_2)$, and $\lambda \in [0, 1]$.
\end{definition}
Essentially, the surplus split factor characterizes collective cooperativeness by revealing how the mixed traffic agents divide the benefits of cooperation. 

\section{Methodology}\label{section:method}

The original concept of collective cooperativeness was developed in a ``virtual'' setting, making it not immediately applicable to real-world scenarios that encompass both longitudinal interactions (e.g., car-following) and lateral interactions (e.g., lane changes and lane selection) between vehicle classes. To bridge this gap and empirically identify and characterize collective cooperativeness, this research seeks to redefine the concept within a more practical and realistic framework. 

\subsection{Representation of collective cooperativeness in multi-space}

\subsubsection{An intuitive example}

\begin{figure}[!ht]
    \centering
    \includegraphics[scale=0.8]{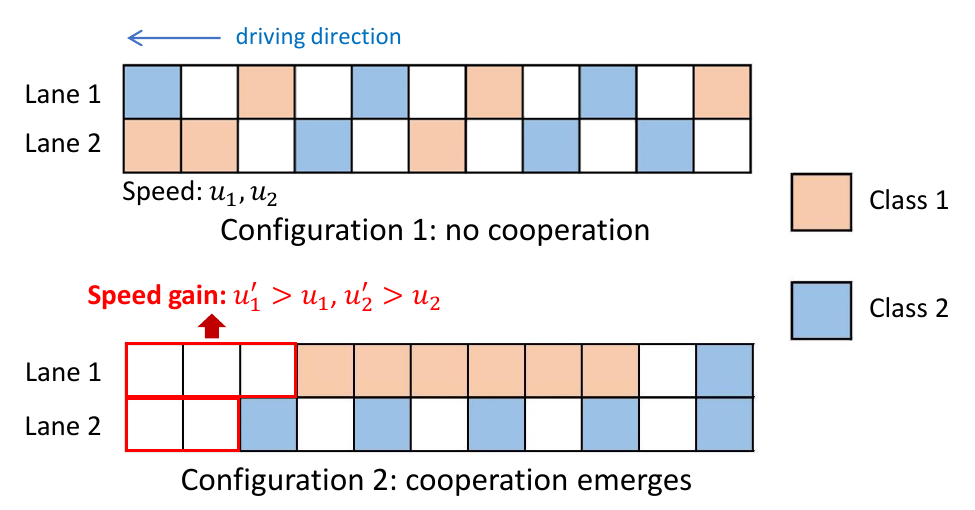}
    \caption{A simple illustration of collective cooperativeness between two classes of traffic agents} 
    \label{fig:configuration}
\end{figure}

We illustrate an intuition of collective cooperativeness in physical world. We consider a simplified scenario involving two classes of vehicles on a two-lane road, as shown in Figure~\ref{fig:configuration}. Each vehicle is depicted as a cell, and the two classes exhibit different desired car-following spacings due to type sensitivity. For instance, class 1 vehicles tend to maintain smaller spacings when following another class 1 vehicle. Essentially, the figure demonstrates that while the number of vehicles for each class are the same, but the configurations of traffic flow are different and lead to different system performance. 

In the first configuration (the upper one), vehicles from both classes are mixed together across the two lanes, traveling without cooperation. The average speed for each class is represented by $u_1$ and $u_2$, respectively.
By contrast, in the second configuration (the lower one), the two classes of vehicles exhibit a certain level of spatial organization. Due to type sensitivity, this spatial organization leads to increased spacing, allowing both classes to experience improved average travel speed compared to when they are fully mixed (i.e., $u_1'>u_1$ and $u_2'>u_2$). In other words, configuration 2 leads to a Pareto improvement over configuration 1, demonstrating the emergence of collective cooperativeness.

\subsubsection{Physical, fluid, and strategic space representation}

\begin{figure}[!ht] 
    \centering
    \includegraphics[width=\textwidth]{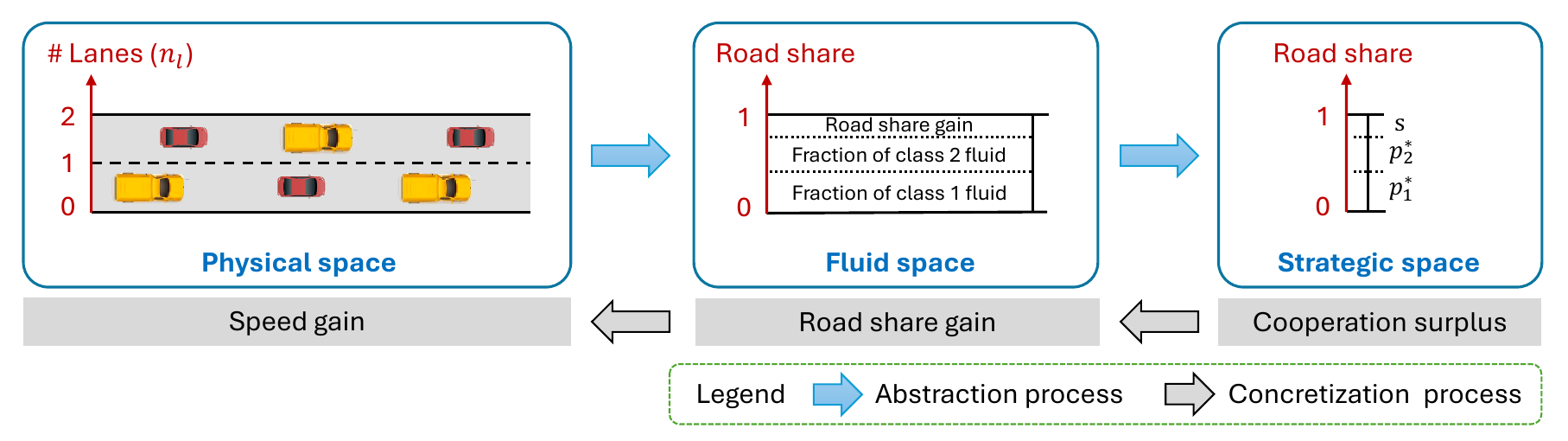}
    \caption{Physical, fluid, and strategic spaces}
    \label{fig:physical_virtual}
\end{figure}

To adopt the theoretical concept in real-world setting, we introduce three different modeling spaces that represent three levels of abstraction, as shown in Figure~\ref{fig:physical_virtual}. Below, we describe each space and the approximations that connect them.

The ``physical space'' represents the actual vehicles on a road segment. In this space, driving agents are distributed across lanes (laterally) and may form platoons (longitudinally), and different spatial configurations lead to different average speeds for each class. The ``fluid space'' abstracts mixed traffic as continuum and the discrete lanes are treated to be infinitely divisible and the total lateral road space normalized to $[0,1]$. The approximation from physical to fluid space is achieved by treating the collection of agents in each class as a traffic stream, akin to classical macroscopic models of mixed traffic. The key difference is that we explicitly consider the lateral dimension of road, which we refer to as ``road share''. In the fluid space, each traffic stream takes a portion of road share, which determines the average speed for the class. The split of road share encapsulates the ability of different classes to take road spaces, which is attributed to their different microscopic characters and behaviors (such as driving aggressiveness and lane preference). 
Then we introduce a ``strategic space'', where the road share split is endogenously determined from agent characteristics. This is achieved by considering the bargaining of different classes on $[0, 1]$. Specifically, each class acts as a self-interested rational player aiming to maximize its road share. Equilibria of the bargaining game correspond to possible splits of road share. Our earlier theoretical model, which develops the collective cooperativeness concept, is applicable to the strategic space.

We further illustrate the connections between these modeling spaces by considering general $\rho_i\,\text{--}\,u_i$ relations. In multi-class traffic, driving agents can have different spatial distributions and interactions, which lead to different $\rho_i\,\text{--}\,u_i$ relation. The concept of road share, denoted by $p_i$,  offers a unified way to formulate  $\rho_i\,\text{--}\,u_i$ relationships~\citep{logghe2008multi, qian2017modeling, li2022equilibrium},
\begin{equation*}\label{eq_beh_metho}
    u_i=u_i^{\text{nominal}}(\rho_i/p_i), \ i=1,2
\end{equation*}
where $u_i$ is the travel speed of class $i$ and $u_i^{\text{nominal}}(\cdot)$ is the nominal speed function for class $i$. In general, $p_i$ is determined by the values of $\rho_1$ and $\rho_2$ (i.e. $p_i=p_i(\rho_1,\rho_2)$). 

The above equation connects the ``physical'' and ``fluid'' spaces depicted in Figure~\ref{fig:physical_virtual}. Here the parameters and functions such as $u_i$, $\rho_i$ and $u_i^{\text{nominal}}(\cdot)$ are observable in the physical space from real-world data. Meanwhile, the road share parameter $p_i$ is defined in the fluid or strategic space. This parameter is not directly observable in the physical world, but may be estimated from the physical-space parameters $u_i$, $\rho_i$, and $u_i^{\text{nominal}}(\cdot)$. Due to the virtual nature of this space, the value of $p_i$ does not necessarily correspond to the number of lanes taken by one class of agents, but rather, it serves as a metric to quantify the efficiency of class $i$.

The primary gap this study seeks to address is characterizing agent interaction within the strategic space using the real-world data. This involves developing an identification framework that could bridge the different spaces. We present the technical details in the subsequent sections.

\subsection{Proposed framework}\label{sec:framework}

We propose a framework to identify and characterize collective cooperativeness in real-world mixed traffic. This framework provides an empirically computable definition of collective cooperativeness in mixed traffic and facilitates the identification of this property from trajectory data. 

\begin{figure}[H] 
    \centering
    \includegraphics[width=\textwidth]{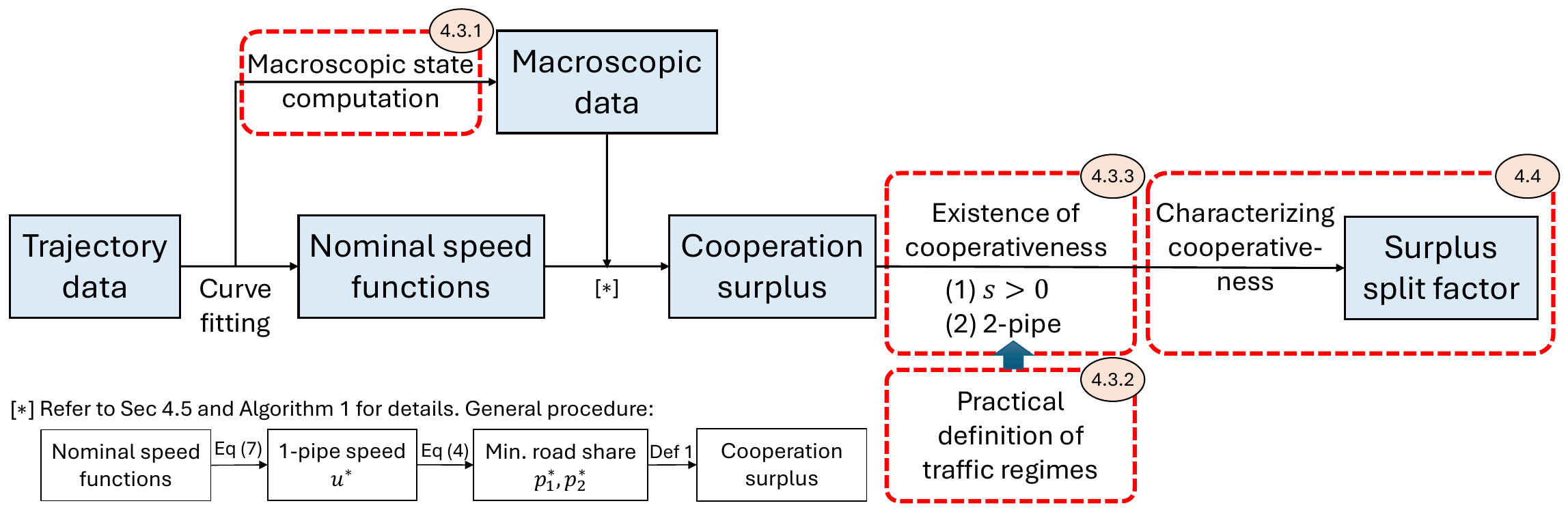}
    \caption{Framework to identify and characterize collective cooperativeness in real-world mixed traffic}
    \label{fig:identification_framework}
\end{figure}

As illustrated in Figure~\ref{fig:identification_framework}, the framework starts with vehicle trajectory data and culminates in the identification and characterization of collective cooperativeness in real-world traffic. This process requires bridging theoretical concepts with measurable phenomena. To accomplish this, the framework incorporates adaptations and refinements to the concept of collective cooperativeness (Section~\ref{sec:adaptaion}). These adaptations include computing macroscopic traffic states from microscopic trajectory data (Section~\ref{sec:micro-macro}), establishing an empirical definition of traffic regimes (Section~\ref{sec:alter_def}), and specifying the conditions for the existence of collective cooperativeness in real-world traffic (Section~\ref{sec:condition_coop}). Once the existence of collective cooperativeness is confirmed, its characterization is detailed in Section~\ref{sec:surplus_split}. Additionally, a summary of the identification process that aligns with the framework is provided in Section~\ref{sec:identification_algo}.

\subsection{Adaptation and refinement of collective cooperativeness concept}\label{sec:adaptaion}

This section seeks to adapt and refine the earlier analytical model by transforming elements that are difficult to measure into quantifiable metrics derived from data. In particular, three main obstacles are addressed.
Firstly, the original notion of collective cooperativeness is defined in a macroscopic equilibrium setting, while real-world traffic consists of individual agents with dynamic and microscopic interactions. Bridging the gap between the micro- and macro-scale measurements is necessary in order to accurately represent traffic states.
Secondly, the fully mixed and fully separate regimes in our original model are a theoretical idealization that do not capture the full range of empirical observations. It is thus necessary to develop a more operable definition of traffic regimes based on direct measurements. 
Lastly, the phenomenon of collective cooperativeness may not be fully observable in real-world traffic. To address this, it is essential to establish an empirical condition that enables the identification of its presence.

In the following, we first establish formulations for computing macroscopic traffic states from microscopic measurements (Section~\ref{sec:micro-macro}). Next, we introduce a practical definition of traffic regimes to enable their empirical characterization (Section~\ref{sec:alter_def}).Finally, leveraging these computations and definitions, we present the empirical conditions for the existence of collective cooperativeness (Section~\ref{sec:condition_coop}).

\subsubsection{Macroscopic state computation}\label{sec:micro-macro}

Traffic flow inherently contains rich information, spanning both microscopic and macroscopic scales. In this study, ``microscopic data'' refers to trajectory data that capture detailed vehicle-level information, such as positions, speeds, and spacings over time. These data are typically collected using high-resolution sensors like video cameras, LiDAR, or GPS. By contrast, the ``macroscopic data'' aggregates traffic information at a system level, summarizing average traffic density and speed over a road segment. These data are usually collected from loop detectors, radar sensors, or computed from microscopic data. In this study, the macroscopic data is computed from trajectory datasets by properly aggregating vehicle-level information, which will be elaborated in this section.

As illustrated in Figure~\ref{fig:identification_framework}, both microscopic data and macroscopic data are integral to the empirical identification of collective cooperativeness. On one hand, microscopic vehicle trajectory data offers detailed car-following events, enabling precise characterization of nominal behaviors that capture vehicle interactions. On the other hand, since collective cooperativeness manifests as a macroscopic phenomenon, its empirical identification necessitates an accurate representation of traffic states at the macroscopic scale. However, unlike microscopic data, which is directly available from the trajectory dataset, macroscopic traffic states need to be carefully computed to ensure an accurate reflection of real-world traffic conditions. This section introduces formulations to derive macroscopic traffic state from microscopic data.

\begin{figure}[!ht] 
    \centering
    \includegraphics[scale=0.5]{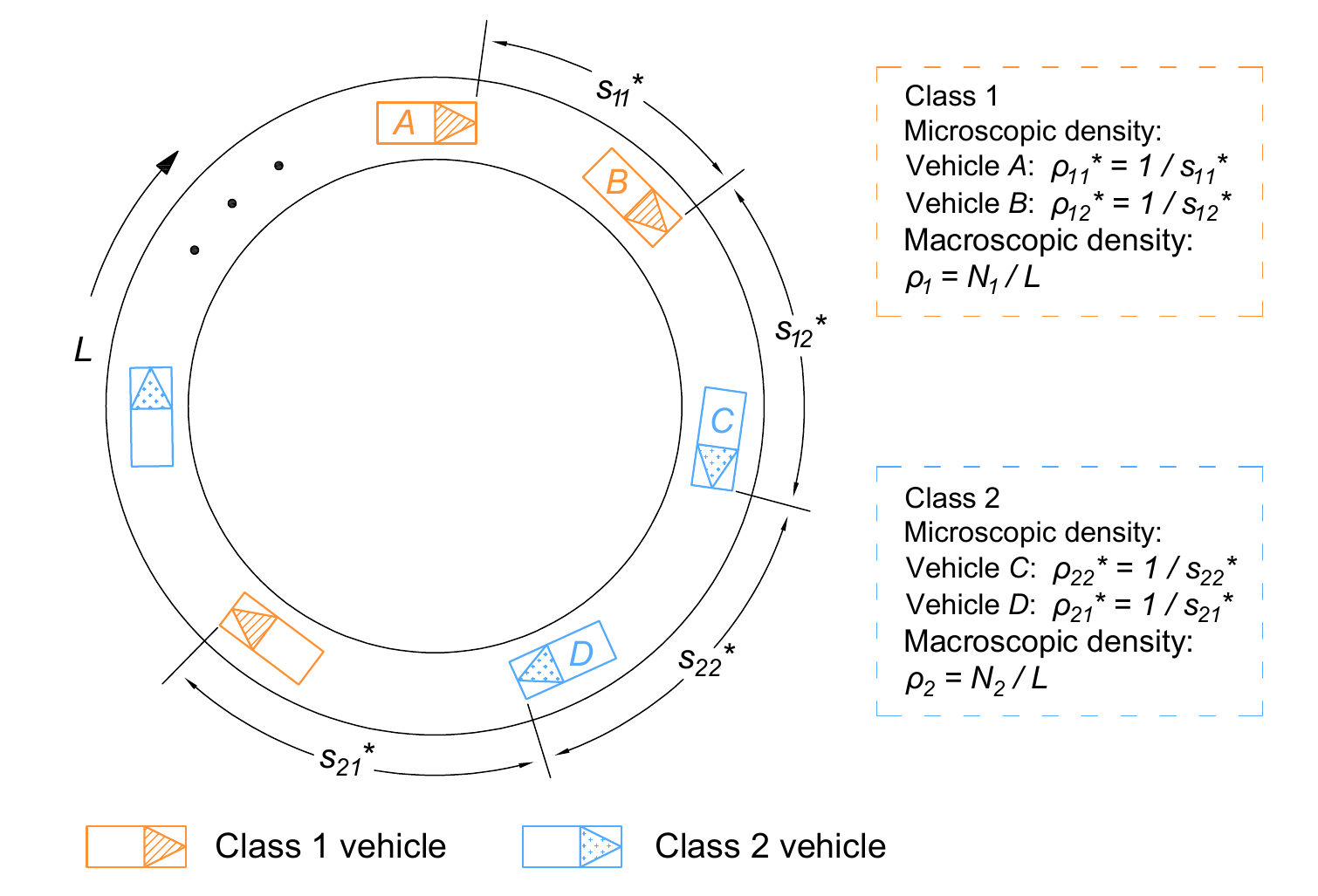}
    \caption{Illustration of microscopic and macroscopic density in ring road with mixed traffic}
    \label{fig:circle}
\end{figure}

We first distinguish between two notions of traffic densities: the microscopic and macroscopic traffic density. We consider a single-lane ring road with a mixture of two classes of vehicles, as depicted in Figure~\ref{fig:circle}. The figure illustrates four car-following scenarios: class 1 following class 1, class 1 following class 2, class 2 following class 1, and class 2 following class 2. Each scenario demonstrates homogeneous car-following behaviors. The total number of class 1 and class 2 vehicles is denoted as $N_1$ and $N_2$, respectively, and the length of the ring road is $L$. 
The micro-level density represents the density as perceived by an individual traffic agent. It captures how a class $i$ agent perceives the overall traffic density based on its immediate surroundings, i.e., the spacing with its class $j$ leading vehicle. In other words, it is defined by the equilibrium spacing between the agent and its leading agent of class $j$, given by $\rho_{ij}^*=1/s_{ij}^*, i,j = 1,2$. In contrast, the macro-level density $\rho_i, i=1,2$, represents a system-level or collective perspective that encompasses all agents, capturing the average density along the studied road segment. Specifically, it describes the average number of class $i$ vehicles per unit length per lane, $\rho_i=N_i/(L n_l)$, where $n_l$ is the number of lanes (which is 1 in this single-lane ring road example).
In homogeneous traffic with only one class of vehicles, the microscopic and macroscopic densities are identical. However, in mixed traffic scenarios, these densities typically differ.

Due to the distinctions between microscopic and macroscopic traffic densities, the macroscopic densities cannot be directly computed from the spacings between vehicles observed in trajectory data. To accurately retrieve macroscopic traffic state data that represents real-world conditions, it is crucial to establish formulations that connect microscopic data to macroscopic states. These formulations are developed by taking snapshots of the studied freeway section.

\begin{figure}[!ht]
    \centering
    \includegraphics[scale=0.6]{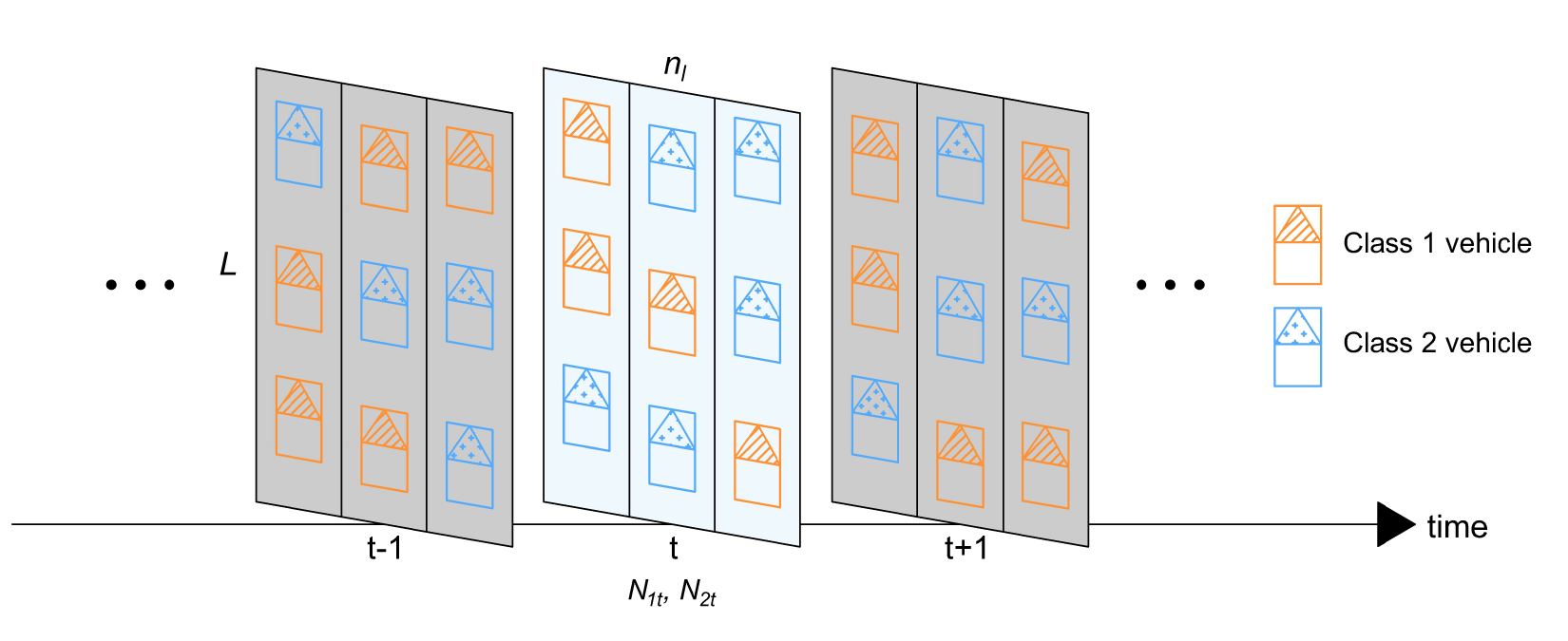}
    \caption{Snapshots of the mixed traffic at different times}
    \label{fig:snapshot}
\end{figure}

As depicted in Figure~\ref{fig:snapshot}, the traffic stream is captured at discrete points in time, with each ``snapshot'' plane (at times $t-1$, $t$, $t+1$, etc.) representing the vehicles present on the studied freeway segment at that specific moment. This approach allows for the calculation of both the average macroscopic density and the average speed for each class of vehicles.

At snapshot $t$, the macroscopic density in each lane is calculated as,
\begin{equation}\label{eq:snapshot_macro_density}
\rho_{it} = \frac{N_{it}}{L n_l}, \ t=1,2,\dots,T; \ i=1,2
\end{equation}
where $N_{it}$ is the number of class $i$ agents in the road segment at time $t$, $n_l$ is the number of lanes, $L$ is the segment length, and $T$ is the total number of snapshots.

The average speed for each class of vehicles at snapshot $t$ is given by,
\begin{equation}\label{eq:snapshot_avg_speed}
    \Tilde{u}_{it} = \frac{1}{N_{it}} \sum_{k=1}^{N_{it}} u_{ikt}, \ i=1,2
\end{equation}
where $u_{ikt}$ is the individual speed of the $k^{th}$ agent from class $i$ at time $t$. The averaged speed $\Tilde{u}_{it}$ reflects a unified movement of each class of agents at time $t$. In fact, the arithmetic mean of speeds formulated in (\ref{eq:snapshot_avg_speed}) is a special case of Edie's generalized definition of traffic speed~\citep{edie1963discussion} when evaluated at a snapshot.

\subsubsection{Practical definition of traffic regimes}\label{sec:alter_def}

We will derive an alternative definition of traffic regimes in this section. This alternative definition is intended to provide a more straightforward and computable approach to measure traffic regimes from empirical data. To differentiate equilibrium traffic regimes from the real-world dataset, the general idea is comparing the empirical speed with the 1-pipe speed predicted from the macroscopic speed function in (\ref{eq:1pipe_speed}). We call it as the regime differentiation rule. Specifically, the rule is written as,
  \begin{equation}
    \text{Traffic regime} = 
    \begin{cases}
    \begin{aligned}
    &\text{2-pipe} & &\text{if $u_{1} \geq u^*$, $u_{2} \geq u^*$ and ``$=$'' cannot hold true simultaneously} \\ 
    &\text{1-pipe} & &\text{if $u_{1} = u_{2} = u^*$} 
    \end{aligned}
    \end{cases}
    \label{eq:regime_empirical}
  \end{equation}
where $u_1$ and $u_2$ are the equilibrium speeds for the class 1 and class 2 vehicles, and $u^*$ is the 1-pipe equilibrium speed computed by applying an appropriate optimization method to solve the condition defined in (\ref{eq:1pipe_speed}). When conditions described in (\ref{eq:regime_empirical}) are not met, we define that Pareto-efficient NE is not attained. 

In below, we show that at Pareto-efficient NE, the equilibrium conditions proposed in (\ref{eq:regime_empirical}) are equivalent to those proposed in our earlier analytical model.

\begin{proposition}[Equivalent condition of Pareto-efficient NE]\label{prop:equiv_theo_empi}
At Pareto-efficient NE, the following equivalence holds: I. The condition of reaching 2-pipe equilibrium $u_{1} \geq u^*$ and $u_{2} \geq u^*$ (the equal sign cannot hold simultaneously) is equivalent to the condition $p_1^*+p_2^* < 1$. II. The condition of reaching 1-pipe equilibrium $u_{1} = u^*$ and $u_{2} = u^*$ is equivalent to the condition $p_1^*+p_2^* \geq 1$.
\end{proposition}

\textit{Proof.}
We prove the equivalence of equilibrium conditions in the 2-pipe and 1-pipe regimes, respectively.

Case I. Equivalent condition of 2-pipe equilibrium.

(1) If $p_1^*+p_2^* < 1$, then $u_1 \geq u^*, u_2 \geq u^*$, and ``$=$'' cannot hold true simultaneously. 

Since $p_1^*+p_2^* < 1$, we have $s=1-p_1^*-p_2^* > 0$. 
By the definition of road share, $p_1=p_1^*+\lambda s, p_2=p_2^*+(1-\lambda)s$, where $\lambda \in [0,1]$. Now, we consider boundary of $\lambda$ case by case. 
When $\lambda \in (0,1)$, the definition of road share gives $p_1 > p_1^*, p_2 > p_2^*$. This leads to $u_1 (\frac{\rho_1}{p_1}) > u_1(\frac{\rho_1}{p_1^*})$ and $u_2(\frac{\rho_2}{p_2}) > u_2(\frac{\rho_2}{p_2^*})$, where the right-hand side of the two inequalities both equals to the 1-pipe speed, $u^*$. Therefore, we have $u_1 > u^*$ and $u_2 > u^*$.
When $\lambda = 0$, the road share that each class of agent takes is $p_1=p_1^*, p_2=p_2^*+s > p_2^*$. In this case, $u_1 (\frac{\rho_1}{p_1}) = u_1(\frac{\rho_1}{p_1^*}) = u^*$ and $u_2(\frac{\rho_2}{p_2}) > u_2(\frac{\rho_2}{p_2^*})=u^*$. This means $u_1=u^*$ and $u_2>u^*$.
By symmetry, when $\lambda=1$, we get $u_1>u^*$ and $u_2=u^*$.
As such, we proved that when $p_1^*+p_2^* < 1$, we have $u_1 \geq u^*, u_2 \geq u^*$, and the ``$=$'' sign never exist simultaneously.

(2) If $u_1 \geq u^*$ and $u_2 \geq u^*$ (``$=$'' do not hold simultaneously), then $p_1^*+p_2^*< 1$. 

For class 1, $u_1 \geq u^*$ means $u_1 (\frac{\rho_1}{p_1}) \geq u^*(\rho_1,\rho_2)$. Then $\frac{\rho_1}{p_1} \leq u_1^{-1}(u^*(\rho_1,\rho_2))$, thus $p_1 \geq \frac{\rho_1}{u_1^{-1}(u^*(\rho_1,\rho_2))} = p_1^*$. Similarly, $p_2 \geq \frac{\rho_2}{u_2^{-1}(u^*(\rho_1,\rho_2))}=p_2^*$. And the equal sign cannot hold true at the same time. 
This means there is at least one class's minimum road share, $p^*$, which is less than its corresponding road share, $p$. Since the total road share equals 1, i.e., $p_1+p_2 = 1$, it follows that $p_1^*+p_2^* < 1$.

In this case, we proved that the 2-pipe conditions of $p_1^*+p_2^* < 1$ and $u_1 \geq u^*$ and $u_2 \geq u^*$ (``$=$'' cannot hold true simultaneously) are equivalent.

Case II. Equivalent condition of 1-pipe equilibrium.

At Pareto-efficient NE, the set of 1-pipe equilibria forms the complement of the set of 2-pipe equilibria. Since the equivalent conditions established for the 2-pipe equilibria (i.e., Case I) hold, the complement of that equivalent conditions remains valid in its complement set (i.e., the 1-pipe equilibrium). This means that the condition for reaching the 1-pipe equilibrium, $u_{1} = u^*$ and $u_{2} = u^*$, is equivalent to the condition $p_1^*+p_2^* \geq 1$. 
The proof is completed.

\begin{figure}[!ht] 
    \centering
    \includegraphics[width=1\textwidth]{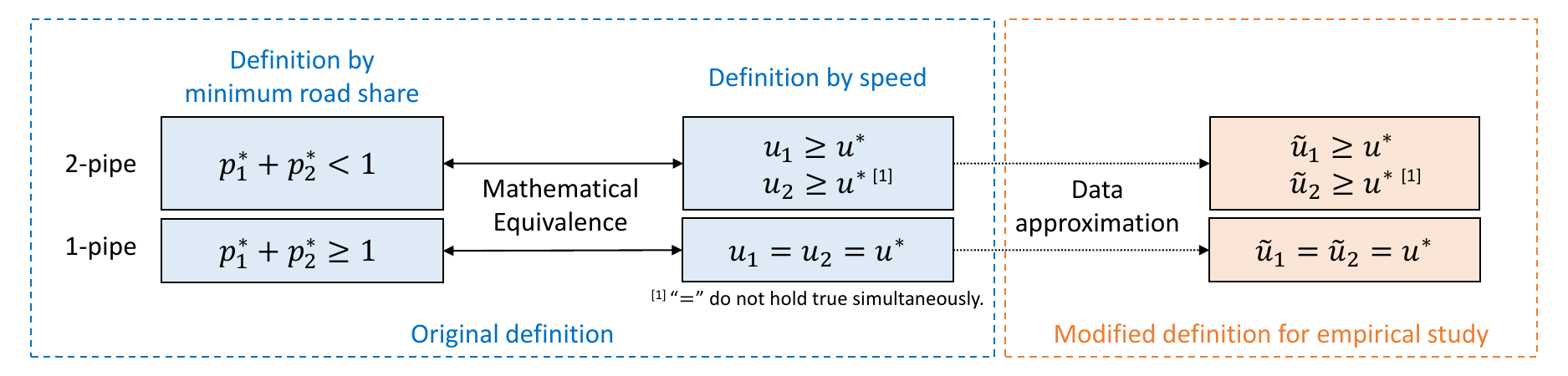}
    \caption{Comparison of the mathematical and empirical definitions of traffic regimes}
    \label{fig:regime_compare}
\end{figure}

While the definitions of traffic regimes are mathematically equivalent, the conditions proposed in (\ref{eq:regime_empirical}) is more convenient for empirical analysis. This condition allows for the substitution of theoretical speeds, $u_1,u_2$, with empirical data, $\Tilde{u}_{1}, \Tilde{u}_{2}$, as illustrated in Figure~\ref{fig:regime_compare}. Therefore, in situations where empirical data are accessible, the proposed equivalent condition offers notable advantages. On one hand, it allows for the substitution of mathematically calculated speeds, $u_i(\cdot)$, with direct measurements, $\Tilde{u}_i$, thereby maximizing the utilization of empirical data and alleviating the reliance on speed assumptions. On the other hand, it simplifies the computational process by eliminating the requirement to solve the inverse nominal speed functions, $u_i^{-1}(\cdot)$. Considering these benefits, we will employ the practical traffic regime definition in (\ref{eq:regime_empirical}) for the empirical analysis.

\subsubsection{Empirical existence condition of collective cooperativeness}\label{sec:condition_coop}

In this section, we focus on identifying the existence of collective cooperativeness. Specifically, we propose conditions for its empirical existence based on the information derived earlier, including the cooperation surplus, which is calculated using macroscopic data from Section~\ref{sec:micro-macro}, and the empirically identified traffic regimes from Section~\ref{sec:alter_def}. The empirical existence condition is described as follows.

\textbf{Empirical existence condition of collective cooperativeness}. 
\textit{We adopt the following conditions for the empirical existence of collective cooperativeness: (1) $s > 0$; and (2) the traffic state operates within the 2-pipe regime. }

The proposed condition is consistent with the game-theoretical model. In fact, this condition corresponds to the subset of Pareto-efficient NE in 2-pipe regime in our analytical model. This condition is suitable for the empirical study, as it decomposes the Pareto-efficient NE into two practical conditions that can be computed from real-world data.

\begin{figure}[ht]
    \centering
    \includegraphics[width=\textwidth]{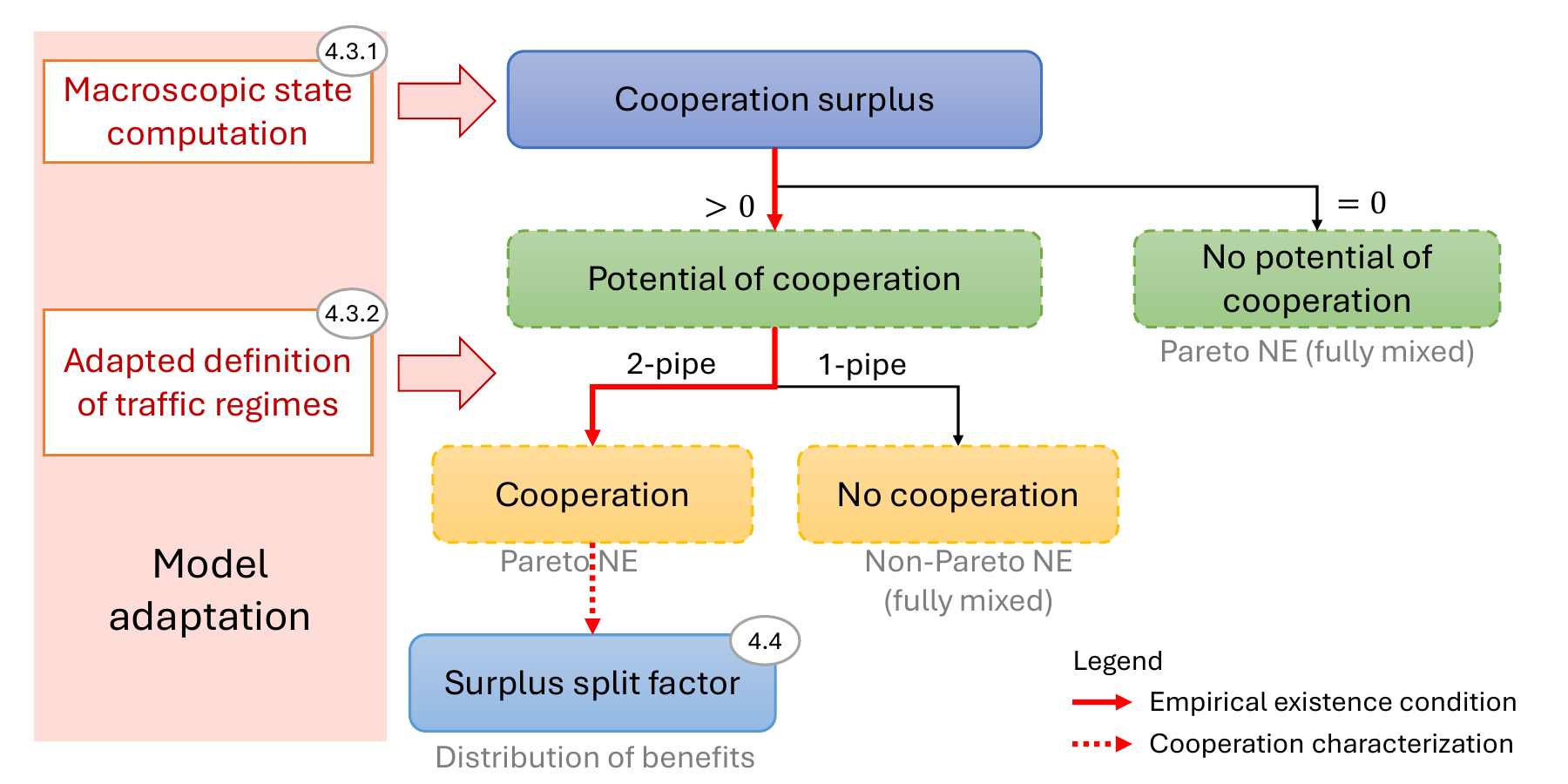}
    \caption{Empirical existence condition and characterization of collective cooperativeness}
    \label{fig:condition_cooperation}
\end{figure}

The workflow for the empirical existence condition of collective cooperativeness is depicted in Figure~\ref{fig:condition_cooperation}. The solid red arrows indicate the conditions under which collective cooperativeness emerges in real-world traffic. Specifically, the condition of $s>0$ suggests the potential for cooperation, while its existence also requires the traffic to operate within the 2-pipe regime. Furthermore, when cooperation is present, the dashed red arrows illustrate the process for characterizing collective cooperativeness. This process is included in the figure for completeness and will be elaborated in the subsequent section. 

\subsection{Identification of surplus split factor} \label{sec:surplus_split}

Now, we seek to measure how the benefits of collective cooperation is allocated among agents, in scenarios where collective cooperation is present. This is achieved by tackling the inverse problem of the empirically adapted model. Essentially, while the equilibria in the bargaining game model are determined by certain parameters, the inverse problem assumes that real-world trajectory data reflect these equilibria and aims to deduce the underlying parameter values. The primary objective is to uncover the hidden surplus split factor from naturalistic driving trajectories, which necessitates examining data both at the micro-level, to capture empirical conditions, and at the macro-level, to align with the macroscopic speed functions.

According to Definition~\ref{def:surplus_split}, the surplus split factor $\lambda$ is the unknown parameter that represents the policy adopted by the two classes of agents to cooperate. The value of it can be determined from different perspectives to reflect different cooperation policies. For instance, when $\lambda$ is independent of $(\rho_1, \rho_2)$, a value of $\lambda=0.5$ indicates an equal split of the cooperation surplus between the two agent classes, while $\lambda=1$ means that class 1 agents greedily take all the cooperation surplus. However, these examples are assumption-based and do not necessarily reflect real-world traffic conditions. In our study, we aim to identify the value of $\lambda$ from empirical observations of mixed traffic flow, which captures the actual interactions between agents. In this case, the value of $\lambda$ depends on the road densities $(\rho_1, \rho_2)$.

We define a loss function to measure the difference between predicted speeds from the macroscopic model and the observed vehicular movement, which is written as,
\begin{equation}\label{eq:lambda_optm_loss}
     L(\rho_1,\rho_2;\lambda) = \frac{1}{T} \sum_{t=1}^T \Bigr[ \sum_{i=1}^2 w_i \lvert \Tilde{u}_{it} - \hat{u}_i(\rho_{1t}, \rho_{2t};\lambda)\rvert \Bigr]^2
\end{equation}
where $\rho_{it}$ and $\Tilde{u}_{it}$ are the macroscopic density and speed computed from the trajectory data by (\ref{eq:snapshot_macro_density}) and (\ref{eq:snapshot_avg_speed}), respectively, $w_i$ are the weights assigned to each class that determines their relative importance in the loss function, $T$ is the total number of snapshots (i.e., the sample size), and $\lambda$ is the surplus split factor. Besides, $\hat{u}_i(\rho_{1t}, \rho_{2t};\lambda)$ is the predicted speed based on the macroscopic speed functions (\ref{eq:payoff}) at the derived equilibria, 
which is expressed as,
\begin{equation}\label{eq:lambda_optm_uhat}
    \hat{u}_i(\rho_{1t}, \rho_{2t};\lambda)=u_i(\frac{\rho_{it}}{p_{it}})=
    \begin{cases}
        u_1(\frac{\rho_{1t}}{p_{1t}^*+\lambda s_t}), & i=1 \\
        u_2(\frac{\rho_{2t}}{p_{2t}^*+(1-\lambda) s_t}), & i=2
    \end{cases}
\end{equation}
Here, $u_i(\cdot)$ represents the nominal speed function for class $i$ agents, $p_{it}$ denotes the road share occupied by each class at time $t$, $p_{it}^*$ indicates the minimum road share for each class, and $s_t$ is the corresponding cooperation surplus. According to Equation (\ref{eq:p1*_p2*}) and Definition~\ref{def:surplus}, we have $p_{it}^*=\frac{\rho_{it}}{u_i^{-1}(u_t^*)}$ and $s_t = 1 - p_{1t}^* - p_{2t}^*$, respectively, where the 1-pipe equilibrium speed $u_t^*$ is solved using (\ref{eq:1pipe_speed}). The surplus split factor $\lambda$ is the only unknown parameter in the loss function. The value of $\lambda$ is estimated by solving 
\begin{equation}\label{eq:lambda_optm}
    \hat{\lambda} = \arg \min_{\lambda} L(\rho_1, \rho_2;\lambda) 
\end{equation}
where $\hat{\lambda}$ is the optimal value of surplus split factor that minimizes the loss. 

We further propose an equity metric to measure the fairness in the allocation of benefits between vehicle classes. To begin with, the estimated surplus split factor is normalized by the percentage of vehicles in each class to eliminate the impact of vehicle population on the benefit allocation. The normalized surplus split factors for each class, denoted as $\tilde{\lambda}_1$ and $\tilde{\lambda}_2$, are defined as,
\begin{equation}\label{eq:lambda_optm_normalize}
    \tilde{\lambda}_1 = \frac{\hat{\lambda}}{P_1}, \ 
    \tilde{\lambda}_2 = \frac{1 - \hat{\lambda}}{P_2}
\end{equation}
where $P_i$ represents the proportion of class $i$ vehicles relative to the total number of vehicles. In calculating $P_i$, we account for the impact of large vehicles on the road by introducing the Passenger Car Equivalent (PCE). Thus, $P_i$ is determined as $P_i = N_i \times PCE_i / \sum_i N_i \times PCE_i$, and it follows $\sum_{i=1}^2 P_i = 1$. We show the following properties hold.

\begin{proposition}[Equity of benefit allocation]\label{prop:lambda_equity} The normalized surplus split factors $\tilde{\lambda}_1$ and $\tilde{\lambda}_2$ have the following properties:
I. If $\tilde{\lambda}_1=\tilde{\lambda}_2$, then it follows that $\tilde{\lambda}_1=\tilde{\lambda}_2=1$. 
II. If $\tilde{\lambda}_1 \neq \tilde{\lambda}_2$, then it follows that $0 \leq \tilde{\lambda}_i < 1$ and $\tilde{\lambda}_j > 1$ for any $i,j\in \{1,2\}$ and $i \neq j$.
\end{proposition}

\textit{Proof.} We prove the conditions I and II respectively as follows.

I. When $\tilde{\lambda}_1=\tilde{\lambda}_2$, by (\ref{eq:lambda_optm_normalize}), we have $\frac{\hat{\lambda}}{P_1}=\frac{1-\hat{\lambda}}{P_2}$. Given that $P_1 + P_2 = 1$, this leads to $\hat{\lambda}=P_1$. In this case, $\tilde{\lambda}_1=\frac{\hat{\lambda}}{P_1}=1$ and $\tilde{\lambda}_2=\frac{1-\hat{\lambda}}{P_2}=\frac{1-P_1}{P_2}=1$. Therefore, when $\tilde{\lambda}_1=\tilde{\lambda}_2$, it must hold that $\tilde{\lambda}_1=\tilde{\lambda}_2=1$.

II. (1) To begin with, we choose $i=1$ and $j=2$ to demonstrate the equivalence between the conditions $0 \leq \tilde{\lambda}_1 < 1$ and $\tilde{\lambda}_2 > 1$. 
(i) Firstly, we show that if $0 \leq \tilde{\lambda}_1 < 1$, then $\tilde{\lambda}_2 > 1$. The lower bound $\tilde{\lambda}_1 \geq 0$ holds because $\tilde{\lambda}_1=\frac{\hat{\lambda}}{P_1}$ and $\hat{\lambda} \geq 0, \ P_1 \geq 0$. Then, given $\tilde{\lambda}_1 < 1$, we have $\frac{\hat{\lambda}}{P_1} < 1$, which implies $\hat{\lambda} < P_1$ and $1-\hat{\lambda} > 1 - P_1$. In this case, $\tilde{\lambda}_2=\frac{1-\hat{\lambda}}{P_2} > \frac{1 - P_1}{P_2}=1$. Thus, if $0 \leq \tilde{\lambda}_1 < 1$, it follows that $\tilde{\lambda}_2 > 1$.
(ii) Secondly, we show that if $\tilde{\lambda}_1>1$, then $0 \leq \tilde{\lambda}_2 < 1$. From $\tilde{\lambda}_1>1$, we have $\frac{\hat{\lambda}}{P_1}>1$, which implies that $\hat{\lambda} > P_1$ and $1-\hat{\lambda} < 1 - P_1$. In this case, $\tilde{\lambda}_2=\frac{1-\hat{\lambda}}{P_2} < \frac{1-P_1}{P_2}=1$. Additionally, since $1-\hat{\lambda} \geq 0$ and $P_2 \geq 0$, the lower bound $\tilde{\lambda}_2 \geq 0$ is satisfied. Therefore, if $\tilde{\lambda}_1>1$, it follows that $0 \leq \tilde{\lambda}_2 < 1$. 
(2) Finally, by symmetry, the equivalent condition also applies when $i=2,\ j=1$. In other words, the equivalence $0 \leq \tilde{\lambda}_2 < 1$ and $\tilde{\lambda}_1 > 1$ holds. This completes the proof.

The relations between $\tilde{\lambda}_1$ and $\tilde{\lambda}_2$ indicate the equity of benefit allocation. If $\tilde{\lambda}_1 = \tilde{\lambda}_2$, the allocation of benefits is equitable, with both values being equal to 1. Otherwise, one of these values will be less than 1 ($0 \leq \tilde{\lambda}_i < 1$), while the other will be greater than 1 ($\tilde{\lambda}_j > 1$, where $j \neq i$). In this scenario, larger values of $\tilde{\lambda}_j$ indicate that this class has greater bargaining power and gains more beyond the average share. 

Subsequently, we introduce an equity metric, denoted as $\zeta$, to quantify the difference in allocated benefits between the two classes. It is defined as,
\begin{equation}\label{eq:equity_metric}
\zeta = \left| \tilde{\lambda}_1 - \tilde{\lambda}_2 \right|
\end{equation}
When $\zeta = 0$, the allocation is perfectly equitable. Conversely, the larger the value of $\zeta$ indicates poorer equity.

\subsection{Summary of identification process}\label{sec:identification_algo}

Based on the proposed framework in Figure~\ref{fig:identification_framework}, we provide a detailed step-by-step procedure for addressing the identification problem, as presented in Algorithm~\ref{algo:identification_procedure}.

In Step~\ref{item:nominal}, microscopic trajectory data is used to identify the nominal speed functions for each class and to estimate the corresponding scaling parameters. Specifically, in Step~\ref{item:nominal}(a), nominal behaviors are determined by selecting a functional form from existing speed-density models, with the model parameters estimated to achieve optimal fit to empirical data. Then, Step~\ref{item:nominal}(b) estimates the scaling parameters, which streamline the computation by reducing the four nominal speed functions ($u_{ij}(\cdot)$) to two nominal speed functions ($u_j(\cdot)$) with two corresponding scaling parameters ($a_j$ and $b_j$).
Step~\ref{item:snapshot} computes macroscopic data from microscopic trajectory data, enabling the analysis of the macroscopic phenomenon of collective cooperativeness.
Then, using the nominal speed functions and macroscopic data, Step~\ref{item:surplus} determines the cooperation surplus. This involves a sequential computation of key parameters in the following order: the 1-pipe speed $u^*$, the minimum road share $p_1^*$, and ultimately the cooperation surplus $s$.
Step~\ref{item:empirical_regime} distinguishes empirical traffic regimes (i.e. 2-pipe or 1-pipe).
Then, based on the computed cooperation surplus and empirical traffic regimes, Step~\ref{item:coop_existence} identifies the existence of collective cooperativeness. Finally, in Step~\ref{item:identification}, collective cooperativeness is characterized by estimating the surplus split factor $\lambda$.

\begin{algorithm}[!h]
\caption{Empirical Identification and Characterization of Collective Cooperativeness}
\label{algo:identification_procedure}
\begin{algorithmic}[1]
    \Statex \textbf{Input:} Vehicle trajectory data of mixed traffic
    \Statex \textbf{Output:} Existence of collective cooperativeness; Surplus split factor
    \Statex \textbf{Steps:}
    
    \begin{list}{\arabic{enumi}.}{\usecounter{enumi}}
       \item \label{item:nominal} Determine nominal behaviors of traffic agents based on the microscopic trajectory data.
       
        \hspace{-1.5em} (a) Identify nominal behaviors of traffic agents, \( u_{ij}(\cdot), i,j\in \{1,2\} \),  by selecting speed-density models for each class and estimating model parameters.
        
        \hspace{-1.5em} (b) Identify scaling parameters \( a_j \) and \( b_j \) based on (\ref{eq:scaling}). This reduces the four nominal speed functions \( u_{ij}(\cdot) \) to two, denoted as \( u_j(\cdot) \).
        
      \item \label{item:snapshot} Compute macroscopic traffic state from trajectory data through snapshots based on (\ref{eq:snapshot_macro_density}) and (\ref{eq:snapshot_avg_speed}). The macroscopic data will be used for the subsequent analysis.
      
      \item \label{item:surplus} Compute cooperation surplus. 

      \hspace{-1.5em} (a) Compute the 1-pipe speed $u^*$ for each pair of macroscopic densities $(\rho_1, \rho_2)$ by (\ref{eq:1pipe_speed}). 
      
      \hspace{-1.5em} (b) Compute the minimum road share $p_i^*$ for each pair of $(\rho_1, \rho_2)$ by (\ref{eq:p1*_p2*}). 
      
      \hspace{-1.5em} (c) Compute the cooperation surplus $s$ according to Definition~\ref{def:surplus} for each pair of  $(p_1^*,p_2^*)$.
      
      \item \label{item:empirical_regime} Determine traffic regimes (1-pipe or 2-pipe) using the regime differentiation rule (\ref{eq:regime_empirical}).
      \item \label{item:coop_existence} Identify the empirical existence of collective cooperativeness based on the proposed condition in Section~\ref{sec:condition_coop}. Output the existence of collective cooperativeness.
      \item \label{item:identification} Formulate the cooperation identification problem by (\ref{eq:lambda_optm_loss}) and (\ref{eq:lambda_optm_uhat}), and estimate the surplus split factor $\lambda$ by (\ref{eq:lambda_optm}). Output the surplus split factor.
  \end{list}
\end{algorithmic}
\end{algorithm}

\section{Case study} \label{section:empirical}

This section presents a case study which applies the proposed framework to real-world data and unveils empirical characteristics of collective cooperativeness. In real world, behaviors of multilane traffic are influenced by factors such as road geometry and lane use rules imposed to different classes of vehicles (though not necessarily enforced). Therefore, the key challenge lies in filtering and preprocessing the data so that the selected data aligns with our modeling assumptions as much as possible.

We start with data preparation in Section~\ref{sec:data_prep}, which takes into consideration the lane use rules and empirical lateral distribution of vehicles to filter and clean the available data. Using the filtered dataset, we identify and characterize collective cooperativeness in Section~\ref{sec:identification}.

\subsection{Data preparation}\label{sec:data_prep}

We utilize the Next Generation Simulation (NGSIM) Vehicle Trajectories and Supporting Data \citep{NGSIM2006} in the case study. The NGSIM dataset contains vehicle trajectories at four locations, including two freeways (I-80 and US-101) and two arterial corridors (Lankershim and Peachtree). Considering the quality and sample size of the data, we select freeway I-80 for analysis, which contains vehicle trajectories at eastbound in Emeryville, CA. The full I-80 dataset covers six freeway lanes, with lane 1 the farthest left lane and lane 6 the farthest right lane, and a high-occupancy vehicle (HOV) lane exists at lane 1. The dataset spans 45 minutes, covering three time intervals: 4:00 - 4:15 p.m., 5:00 - 5:15 p.m., and 5:15 - 5:30 p.m. Three vehicle types, namely, cars, trucks, and motorcycles, are included. We recognize that the NGSIM dataset does not cover all the traffic regimes outlined in our theory, as it mainly covers the congestion regime. Other public datasets also face similar limitations. Our objective is to showcase the proposed identification methodology to traffic in the observed regimes and analyze corresponding characteristics.

\subsubsection{Lane group selection}\label{sec:data_select}

Freeways consist of multiple lanes. Usually, lane policies can be implicitly or explicitly imposed (e.g. restriction of trucks to use left lanes) and varying traffic dynamics can be exhibited across different lane groups~\citep{duret2012lane, hamdar2009life}. Therefore, it is necessary to carefully select a suitable set of lanes, so that the impact of lane use policies is minimized. To this end, we first classify the lanes on I-80 into three different groups based on their respective functionalities and then consider the impact of lane policy on each lane group. 

Specifically, we categorize the studied I-80 segment into three lane groups, namely, the HOV lane group, the ramp and slow vehicle lane group, and the middle lane group, each detailed below:

\begin{itemize}
    \item The HOV lane group is exclusively designated for high occupancy vehicles. It comprises the left-most lane on I-80.
    \item The ramp and slow vehicle lane group comprises lanes designated for entering and exiting the freeway, as well as for slower-moving vehicles. On I-80, this group encompasses the two right-most lanes, lanes 5-6. These lanes usually contain a large proportion of the entering and exiting traffic. Besides, slower-moving vehicles, such as trucks with three or more axles, are mandated to drive in this lane group with speed below 55 mph, according to the California Vehicle Code (CVC) regulations~\citep{california_vehicle_code}.    
    \item The middle lane group does not have regulatory constraints, and we assume that the majority of through traffic utilize this lane group. It includes lanes 2-4 on I-80.
\end{itemize}

Among these lane groups, the HOV lane group is excluded from the analysis due to its restricted access and concentration of passenger cars. The ramp and slow vehicle lane group is unsuitable because of two main factors. On one hand, some vehicles in this lane group have limited lane choices due to the imposed regulation by CVC. On the other hand, vehicles may employ lane changes for exiting or entering the freeway rather than gaining speed. This is indicated by the higher number of lane changes (LCs) in Table~\ref{table:lane_group}. This means vehicles' utilities are not fully described by speed gain, as assumed in our model. 

We choose the middle lane group for the ensuing analysis, based on two rationales. First, vehicles in this lane group possess a simpler utility function focused primarily on speed gain, representing self-interested behavior as defined in this study. Second, lanes in this group are interchangeable due to their similar functionality, which suggests that drivers' lane choices are driven by their endogenous motive of gaining speed.

In addition, we restrict focus on the mixed traffic flow of cars (referred to as Class 1) and trucks (referred to as Class 2). This leads to four different car-following types, namely, car following car (car-car), car following truck (car-truck), truck following truck (truck-truck), and truck following car (truck-car).

\begin{table}[!ht]
    \centering
    \begin{threeparttable}
        \caption{Statistics among lane groups on I-80}
        \label{table:lane_group}
        \begin{tabular}{cccc}
            \toprule
            Lane group & Lanes & Avg. LCs per lane & \% of trucks\tnote{*} \\
            \midrule
            HOV lane group & Lane 1 & - & 3.66  \\
            Middle lane group & Lanes 2-4 & 0.69 & 1.85  \\
            Ramp and slow vehicle lane group & Lanes 5-6 & 1.00 & 3.44\\
            \bottomrule
        \end{tabular}
        \begin{tablenotes}
            \small
            \item[*] While CVC regulations indicate that trucks should utilize the ramp and slow vehicle lane group on I-80, it remains uncertain regarding whether the exact definition of trucks in the NGSIM dataset aligns with the definition by CVC. By reviewing the NGSIM videos, we noticed trucks in other lanes besides lanes 5-6. Thus, we assume that trucks observed in other lane groups are legally permitted to operate within these lanes.
        \end{tablenotes}
    \end{threeparttable}
\end{table}

\subsubsection{Data filtering and cleaning}\label{sec:data_filter}

\begin{figure}[!ht]
    \centering
    \includegraphics[scale=0.45]{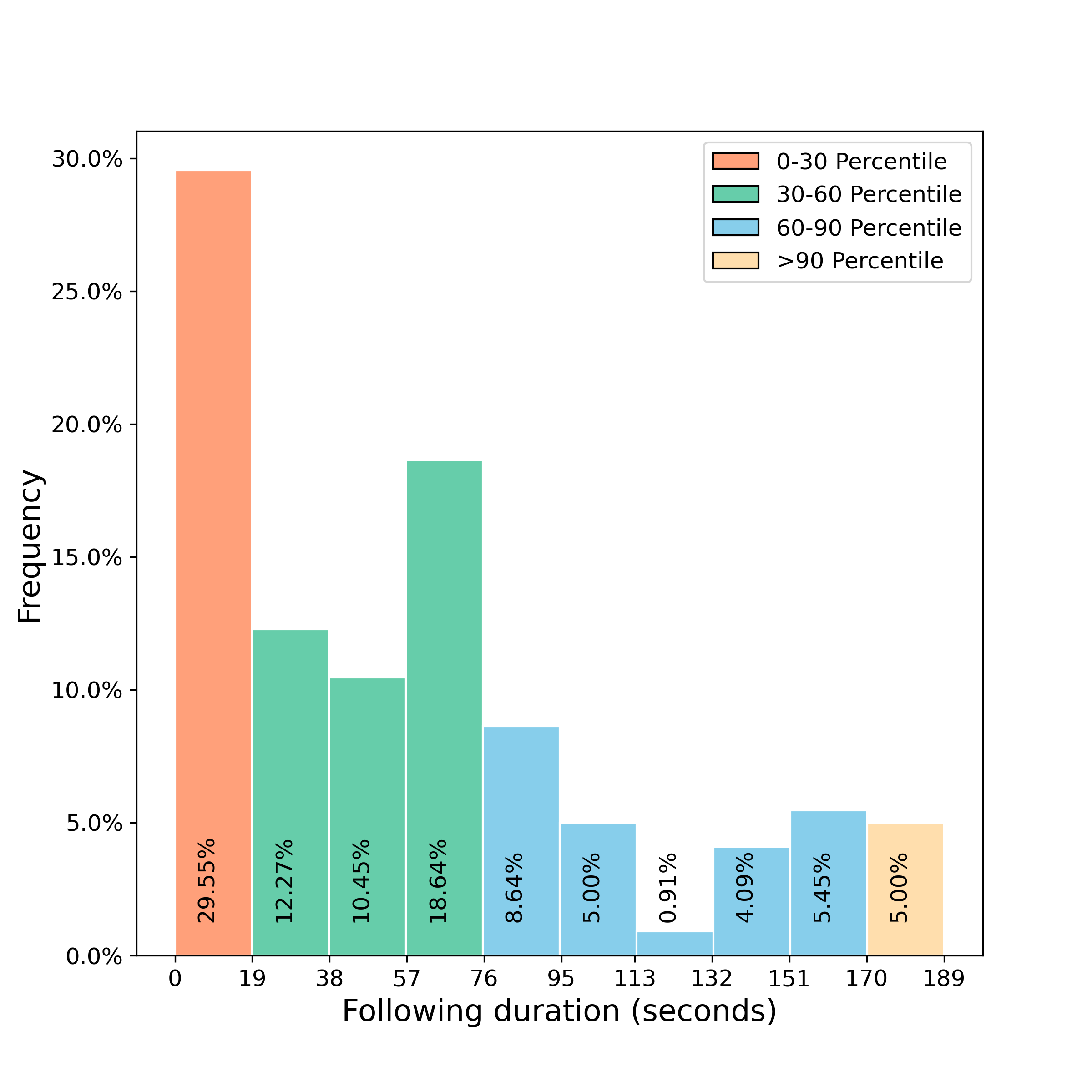}
    \caption{Car-following duration distribution at I-80}
    \label{fig:follow_duration}
\end{figure}

To reduce noise in the trajectory data for the middle lane group and retrieve equilibrium conditions, we apply data filtering based on the following criteria:
\begin{itemize}
    \item The maximum acceleration and deceleration rates are set to 1 $m/s^{2}$ to eliminate abrupt speed changes and potential data errors. In this process, only the abnormal portions of the trajectory that exceed these limits are discarded, ensuring that as much trajectory information as possible is retained.
    \item The car-following duration should be at least 60 seconds to ensure stable car-following behavior while providing sufficient data for analysis, as depicted in Figure~\ref{fig:follow_duration}.
    \item A 10-second cut-off window is applied at the beginning and end of each car-following process to exclude trajectory records that are prone to containing external disturbances.
    \item While lane changes can occur in this lane group, each car-following process excludes lane changes. This ensures that each car-following behavior takes place in a same lane.
\end{itemize}

Besides, it is assumed that all vehicles in the identified leader-follower pairs are interacting. This assumption is supported by two rationales. First, the NGSIM dataset explicitly identifies leader-follower pairs, and we adhered to this definition to maintain consistency. Second, our preliminary analysis indicates that vehicles in the studied dataset typically maintain small spacings due to dense traffic conditions, with an average spacing of 14.81 meters and a 75th percentile spacing of 17.02 meters, suggesting a high likelihood of interaction between vehicles. We also note that in other datasets where larger spacings may occur, defining a threshold for car-following spacing could be helpful to ensure that identified pairs accurately reflect vehicle interactions.

Moreover, to further ensure the correctness of the car-following trajectories in the middle lane group, we examined the distance-time diagrams for each car-following pair, manually excluding any records depicting vehicle collisions or containing obvious errors. 

\begin{figure}[ht] 
    \centering
    \includegraphics[scale=0.8]{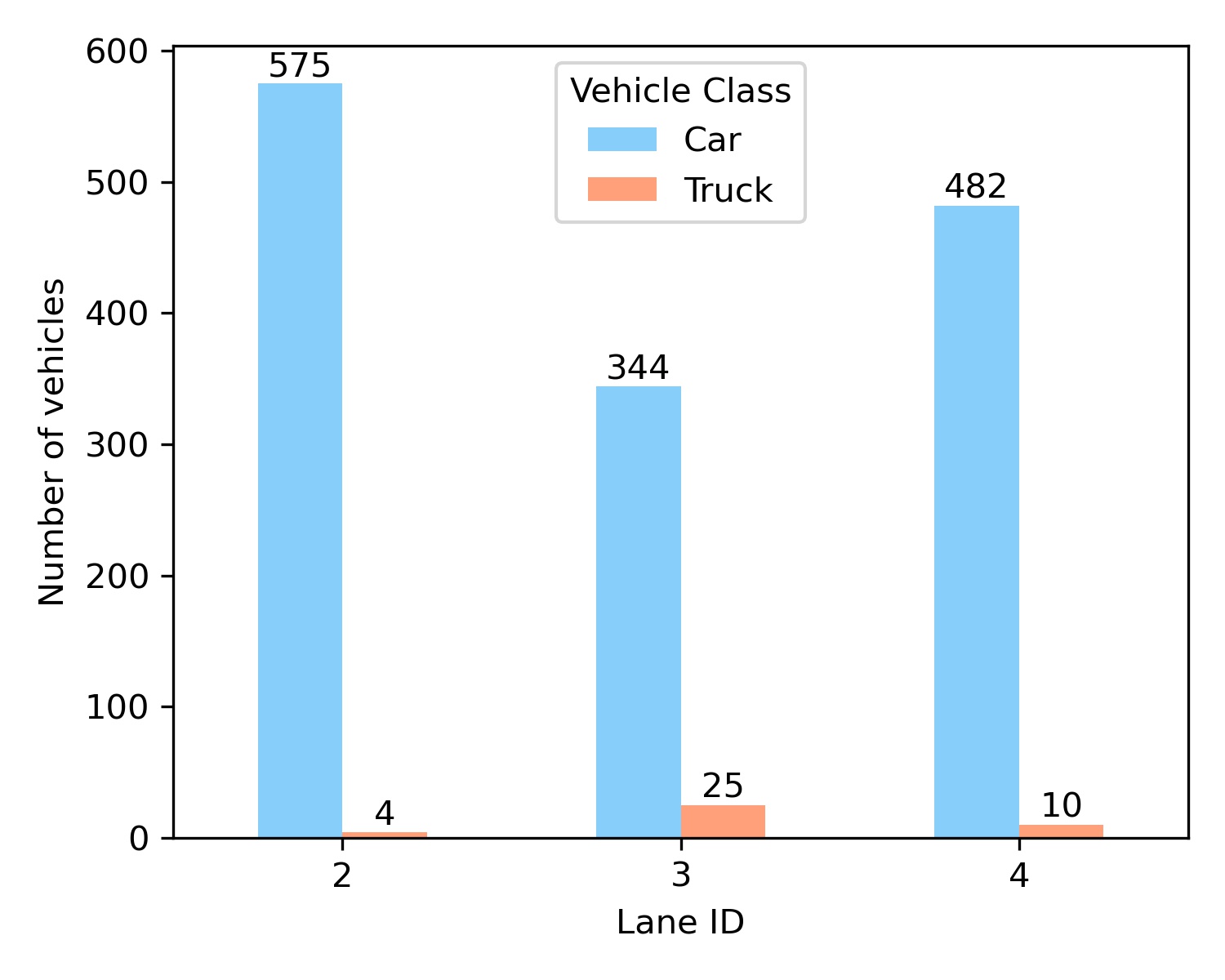}
    \caption{Lane distribution on lanes 2-4 at I-80}
    \label{fig:lane_distribution}
\end{figure}

The resultant dataset comprises 1,907 car-following pairs, including 1,830 car-car, 35 car-truck, 3 truck-truck, and 39 truck-car pairs. It encompasses a total of 965,438 trajectory records, with 938,571 car-car, 21,558 car-truck, 1,297 truck-truck, and 4,012 truck-car trajectories. This final dataset spans a segment of 544 meters along the studied I-80 segment. The distribution of vehicle classes across lanes are shown in Figure~\ref{fig:lane_distribution}. 

\subsection{Identification results}\label{sec:identification}

In this section, we analyze the processed dataset to validate and characterize collective cooperativeness in real-world traffic scenarios. Following the identification framework in Figure~\ref{fig:identification_framework}, the microscopic trajectory data is utilized to identify nominal speed functions (Section~\ref{section:empirical_nominal}), while the macroscopic data is employed to identify the cooperation surplus and determine the empirical existence of collective cooperativeness (Section~\ref{section:empirical_surplus}), and characterize cooperativeness by estimating the surplus split factor (Section~\ref{section:empirical_split_factor}). We follow the detailed steps in Algorithm~\ref{algo:identification_procedure} for the identification procedure.

\subsubsection{Nominal speed functions and scaling parameters}\label{section:empirical_nominal}

As depicted in Figure~\ref{fig:identification_framework}, we start by determining the nominal behavior for each vehicle class and their respective scaling parameters using microscopic trajectory data. The reason for using micro-scale data for this task is that the microscopic trajectory data offers a wealth of car-following events. Such car-following information have been widely utilized to construct traffic flow models, such as car-following model~\citep{kesting2008calibrating}, speed-spacing model~\citep{duret2008estimating}, etc. In our case, this granularity enables a more precise identification of nominal speed-density functions that encompasses vehicle interactions, and facilitates the quantification of type-dependencies between vehicle classes.

After an initial investigation and selection process, we chose to utilize the logistic speed-density model to represent the nominal behavior for cars~\citep{wang2011logistic}, and the Underwood speed-density model for trucks~\citep{underwood1961speed}. Both models offer a reasonable fit with the studied dataset and are chosen for their mathematical clarity and the ease of understanding their parameters. In the following, we denote the speed-density function for class 1 as $f(\cdot)$, and for class 2 as $g(\cdot)$.

The five-parameter logistic speed-density model is written as,
  \begin{equation}
    f(\rho)=u_b + \frac{u_f-u_b}{\left(1+\exp(\frac{\rho-\rho_c}{\theta_1})\right)^{\theta_2}}
\end{equation}
where $u_f$ is the free-flow speed, $u_b$ is the average travel speed during the trip, $\rho_c$ denotes the critical density at which the traffic transitions from free-flow to congested flow. The scale parameter $\theta_1$ determines the elongation of the curve across the density range, and parameter $\theta_2$ influences how lopsided the curve is.

The Underwood model, which has fewer parameters compared to the logistic model, takes the form of,
  \begin{equation}
    g(\rho)=u_f \exp(-\frac{\rho}{\rho_c})
  \end{equation}

The model parameters for each car-following type are estimated using the Improved Stochastic Ranking Evolution Strategy (ISRES) algorithm \citep{runarsson2005search}. The ISRES algorithm aims to minimize a nonlinear objective function by employing log-normal step-size update and exponential smoothing techniques. 
In our study, the objective function is defined as the $L_1$-norm loss function $l(\theta)=\sum_{k=1}^N |u_k - h(\rho_k;\theta)|$, where $u_k$ denotes the empirical speed for the $k$th observation, $N$ is the sample size, and $h(\rho_k,\theta)$ represents any nominal speed-density model with parameter set $\theta$. 
In our case, when using the logistic model, $h(\cdot)$ corresponds to $f(\cdot)$, and the parameter set is $\theta=(u_b,u_f,\rho_c,\theta_1,\theta_2)$; when using the Underwood model, $h(\cdot)$ corresponds to $g(\cdot)$, and the parameter set is $\theta=(u_f,\rho_c)$. Constraints on these parameters are applied to maintain the models' physical realism. 
Besides, each car-following type is analyzed separately to estimate parameters of the corresponding speed functions: $f_{11}(\cdot)$ for car-car interactions, $f_{12}(\cdot)$ for car-truck interactions, $g_{22}(\cdot)$ for truck-truck interactions, and $g_{21}(\cdot)$ for truck-car interactions. And the speed and density data at the microscopic level are utilized to estimate these parameters. Moreover, the dataset is divided into a 70\% training set and a 30\% testing set, where the parameters are estimated on the training set, while the testing set is used to evaluate out-of-sample predictive accuracy.

We calculate the mean absolute error (MAE) to assess the estimation accuracy. For each car-following type, the MAE is determined by comparing the observed speeds from the corresponding trajectory data with the predicted speeds from the respective speed-density model. It is expressed as,
\begin{equation}\label{eq:MAE}
    MAE_{ij} = \frac{1}{N_{ij}} \sum_{k=1}^{N_{ij}} |u_k - \hat{u}_k|, \ i,j \in \{1,2\}
\end{equation}
where $MAE_{ij}$ represents the mean absolute error for the case of class $i$ following class $j$, $N_{ij}$ is the sample size for that car-following type, $u_k$ is the empirically observed speed from the corresponding trajectory data. The term $\hat{u}_k$ represents the speed predicted by the speed-density model based on the estimated parameters.

Table~\ref{tab:logistic_underwood} presents the parameter estimates obtained for the logistic and Underwood models for various types of car-following behavior, as well as the MAEs computed by (\ref{eq:MAE}).

\begin{table}[h]
    \centering
    \caption{Optimized parameters for the logistic and Underwood speed-density model}\label{tab:logistic_underwood}
    \begin{tabular}{ccccccccc}
    \toprule
    Model & Ego & Leader & $u_b$ (mph) & $u_f$ (mph) & $\rho_c$ (vpm) & $\theta_1$ & $\theta_2$ & MAE (mph) \\ 
     \midrule
     \multirow{2}{*}{logistic} & \multirow{2}{*}{car} & car & 7.93 & 73.55 & 20.40 & 8.0387	& 0.2309 & 4.18 \\
        & & truck & 6.63 & 67.87 & 22.73 & 3.4839 & 0.2673 & 4.69 \\ 
     \cline{3-9}
     \multirow{2}{*}{Underwood} & \multirow{2}{*}{truck} & truck & - & 42.55 & 41.74 & - & - & 2.46 \\ 
        & & car & - & 60.10 & 79.49 & - & - & 7.81 \\ 
    \bottomrule
    \end{tabular}
    \label{table:param}
\end{table}

Utilizing the selected speed-density models and estimated model parameters in Table~\ref{tab:logistic_underwood}, we now verify the scaling property for each vehicle class. The speed functions for car-car, $f_{11}(\cdot)$, and truck-truck, $g_{22}(\cdot)$, are adopted as the nominal speed functions, represented by $f(\cdot)$ and $g(\cdot)$, respectively. This results in $a_1=1, b_2=1$. By applying the scaling property described in (\ref{eq:scaling}), the speed functions for car-truck and truck-car are respectively expressed as,
\begin{equation}\label{eq:scaled_model}
f_{12}(\rho)=f(\rho/a_2),
\ g_{21}(\rho)=g(\rho/b_1)
\end{equation}
where $a_2$ is the scaling parameter for car-truck interactions, and $b_1$ is the one for truck-car interactions.

Our goal is to estimate the scaling parameters $a_2$ and $b_1$ by reducing the $L_1$-norm error between the observed speeds and predicted speeds. For each scenario (i.e., car-truck or truck-car), the $L_1$-norm loss is expressed as $l(q)=\sum_{k=1}^{N} |u_k-h(\rho_k; q)|$. Here, $u_k$ represents the empirical speeds obtained from the respective trajectory data, $N$ is the corresponding sample size, and $q$ denotes the scaling parameter to be estimated (i.e., $a_2$ or $b_1$). Additionally, $h(\rho_k; q)$ represents the model predicted speed given the scaling parameter $q$ based on (\ref{eq:scaled_model}). The ISRES algorithm is employed to estimate these parameters. 

We use the MAE metric proposed in (\ref{eq:MAE}) to evaluate the estimation accuracy. In particular, $MAE_{12}$ and $MAE_{21}$ are calculated by comparing the observed speeds in car‐truck or truck‐car trajectories against the predicted speeds from $f(\rho/a_2)$ or $g(\rho/b_1)$, respectively.
Table~\ref{table:scaling} presents the resulting scaling parameters and the corresponding MAEs on the test dataset. A scaling parameter value lower than 1 implies a larger headway at the same speed compared to the reference headway. Consequently, the result suggests that cars generally maintain a larger headway when following trucks than when following other cars, whereas trucks tend to follow cars more closely than they do other trucks. We postulate that this is because trucks have a higher vantage point and better visibility, enabling them to see farther ahead and down the road. 

\begin{table}[h]
    \centering
    \caption{Estimated scaling parameters}
    \label{table:scaling}
    \begin{threeparttable}
    \begin{tabular}{ccc}
    \toprule
     Nominal speed function & Scaling parameter & MAE (mph)$^1$ \\ 
     \midrule
     logistic & 0.4528 & 4.58 \\ 
     Underwood & 2.5996	& 6.46 \\ 
     \bottomrule
    \end{tabular}
     \begin{tablenotes}
        \small
        \item[1] The MAEs in this table exhibit a slight reduction compared to the corresponding MAEs in Table~\ref{tab:logistic_underwood}. This arises from the difference in functional forms.
    \end{tablenotes}
    \end{threeparttable}
\end{table}

\begin{figure}[!ht]
\centering
\begin{subfigure}[t]{.45\textwidth}
  \centering
  \includegraphics[width=\linewidth]{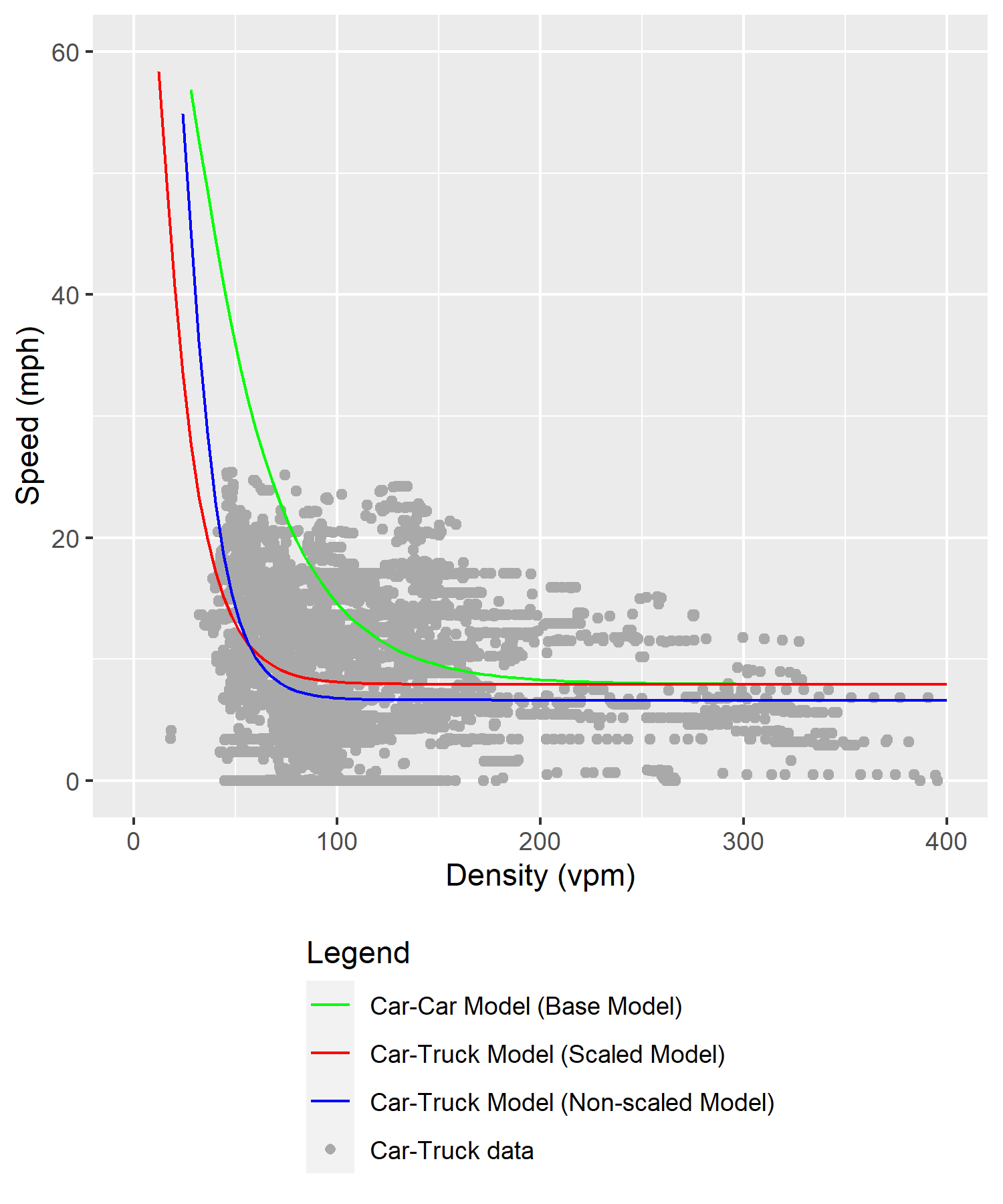}
  \caption{Cars (Class 1)}
  \label{fig:scale_car}
\end{subfigure}
\begin{subfigure}[t]{.45\textwidth}
  \centering
  \includegraphics[width=\linewidth]{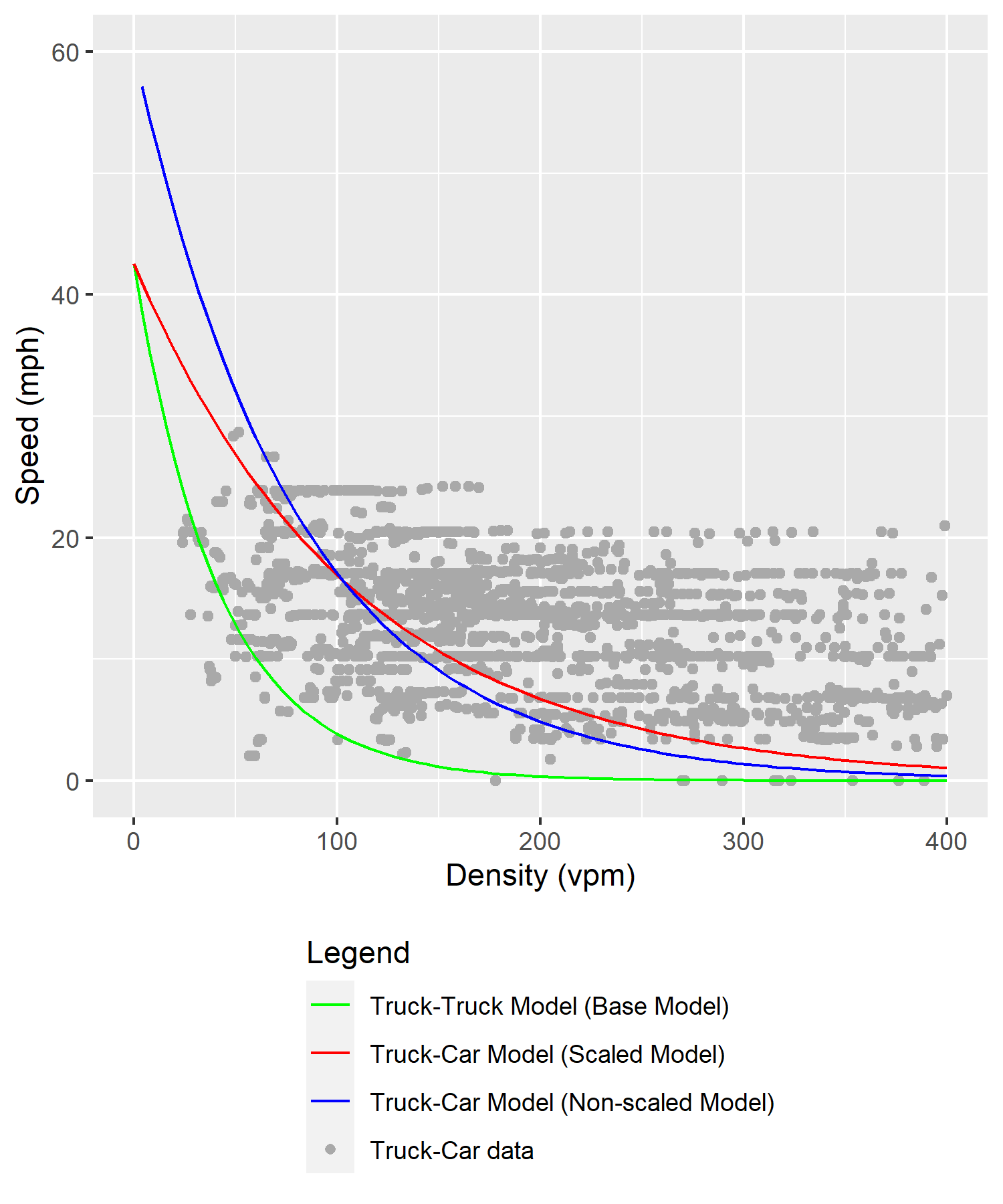}
  \caption{Trucks (Class 2)}
  \label{fig:scale_truck}
\end{subfigure}
\vspace{0.2cm}
\caption{Comparison of base, scaled, and non-scaled logistic (left) and Underwood (right) models} 
\label{fig:scaledmodel_all}
\end{figure}

Figure~\ref{fig:scaledmodel_all} presents a comparison of the fitted speed functions. In the figure, the data points for car-truck and truck-car interactions are visualized respectively in (a) and (b). The blue curve represents the original speed-density model fitted to the corresponding data points (i.e., $f_{12}(\rho)$ for car-truck scenario and $g_{21}(\rho)$ for the truck-car scenario, with estimated parameters listed in Table~\ref{tab:logistic_underwood}). The green curve represents the nominal speed-density models $f(\rho)$ and $g(\rho)$, specifically referring to $f_{11}(\rho)$ and $g_{22}(\rho)$, using parameters from Table~\ref{tab:logistic_underwood}. Additionally, the red curve depicts the scaled models of $f_{12}(\rho)$ and $g_{21}(\rho)$ based on (\ref{eq:scaled_model}), with estimated parameters provided in Table~\ref{table:scaling}. 
As can be seen in the figures, in both logistic and Underwood models, the blue and red curves demonstrate a reasonable approximation to each other, distinct from the reference speed function (green curves). This observation suggests that substituting the car-truck and truck-car models with a nominal speed function featuring a scaling parameter could be a viable approach. This substantiates the applicability of the scaling property. In addition, it is reasonable that the green curves deviate from the data points since they correspond to different car-following types; they are included here for comparison with the scaled model (the red curve). Moreover, given the dispersion of the data, we believe these models offer a reasonable degree of fit. 

\subsubsection{Empirical identification of cooperation surplus}\label{section:empirical_surplus}

This section aims to measure the empirical values of cooperation surplus, determine the existence of collective cooperativeness, and quantify the likelihood of its emergence.

To begin with, given that collective cooperativeness manifests as a macroscopic phenomenon, we compute macroscopic traffic state data from the microscopic trajectory data and employ the macroscopic data to facilitate our investigation. By taking snapshots every 0.5 seconds, we compute the macro-level density and the mean speed at each interval, as described in (\ref{eq:snapshot_macro_density}) and (\ref{eq:snapshot_avg_speed}). This process corresponds to Step~\ref{item:snapshot} in Algorithm~\ref{algo:identification_procedure}.

In the subsequent paragraphs, we will perform the computation of key parameters, including the 1-pipe speed and the empirical cooperation surplus, and the identification of the empirical existence of collective cooperativeness. These computations and identifications correspond to Steps~\ref{item:surplus} to~\ref{item:coop_existence} of Algorithm~\ref{algo:identification_procedure}. Furthermore, we will look deeper into the characteristics and implications of these computed values.

\noindent
\textbf{1-pipe speed}

As outlined in Step~\ref{item:surplus}(a) of the algorithm, the 1-pipe speed $u^*$ is computed using the implicit function defined in (\ref{eq:1pipe_speed}). In this equation, $u^*$ is the only unknown parameter to be determined, while the nominal speed functions $u_i(\cdot)$ and the scaling parameters $a_j$ and $b_j$ are obtained from Step~\ref{item:nominal}, and the macroscopic traffic densities $\rho_1$ and $\rho_2$ are obtained from Step~\ref{item:snapshot}. To solve for $u^*$, we employ Brent's method~\citep{brent1971algorithm}, a numerical optimization algorithm used for solving univariate minimization problems. This approach allows the computation of $u^*$ for each pair of macroscopic density values $(\rho_1, \rho_2)$.

\begin{figure}[!ht]
    \centering
    \includegraphics[scale=0.45]{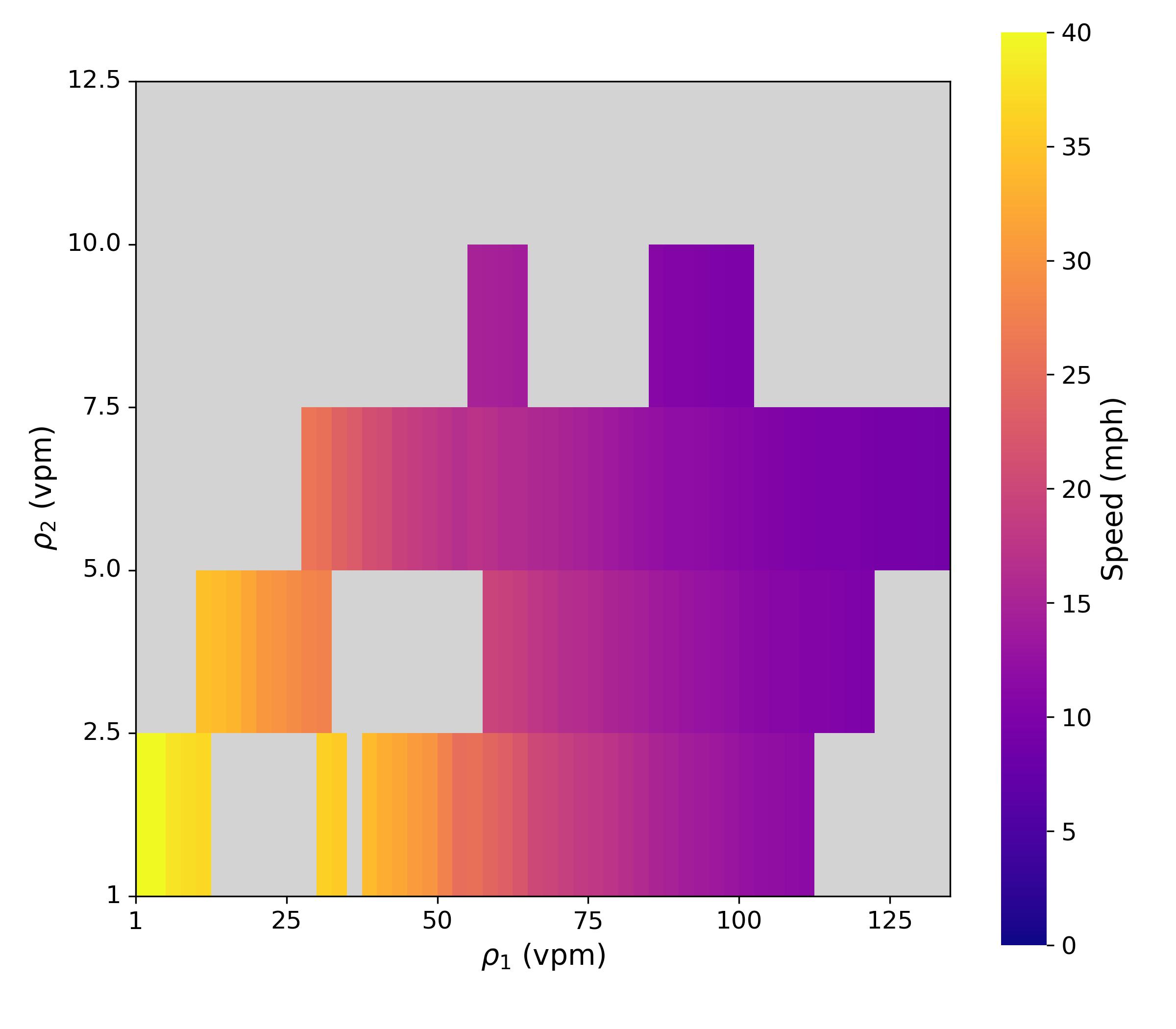}
    \caption{Heatmap of 1-pipe speed in NGSIM data. The gray area results from a lack of data. Besides, due to the limited number of trucks in the studied segment, the interval in the y-axis ($\rho_2$) is much lower than the interval in the x-axis ($\rho_1$).}
    \label{fig:spd1p_headmap}
\end{figure}

The heatmap of the computed one-pipe speeds is shown in Figure~\ref{fig:spd1p_headmap}. It is observed that the one-pipe speeds are generally higher at lower traffic densities and diminish with increasing density. This is consistent with practical scenarios where vehicles travel faster when there are fewer vehicles on road.

\noindent
\textbf{Cooperation surplus}

Following Step~\ref{item:surplus}(b) in Algorithm~\ref{algo:identification_procedure}, the minimum road shares $p_1^*$ and $p_2^*$ can be determined using the computed values of $\rho_1$, $\rho_2$, and $u^*$. Subsequently, in Step~\ref{item:surplus}(c), the cooperation surplus can be easily calculated based on Definition~\ref{def:surplus}.

\begin{figure}[!ht]
    \centering
    \includegraphics[scale=0.7]{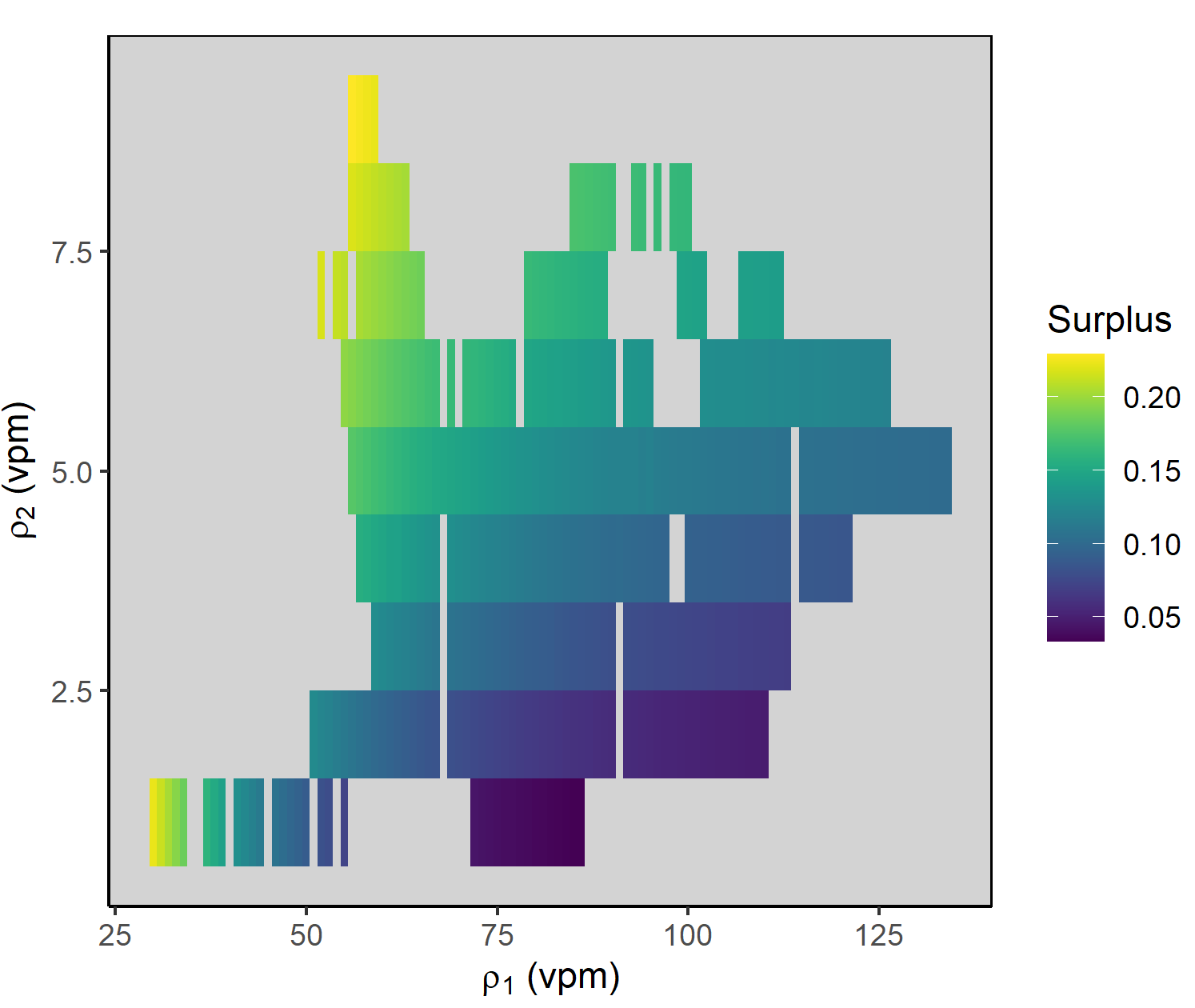}
    \caption{Surplus heat map over class 1 and class 2 macroscopic densities}
    \label{fig:surplus_headmap}
\end{figure}

Figure~\ref{fig:surplus_headmap} presents the heatmap depicting the values of cooperation surplus across different macroscopic densities. The surplus values are positive, implying the potential of cooperation within the analyzed freeway segment, as illustrated in Figure~\ref{fig:condition_cooperation}. Besides, the surplus values tend to be higher at lower densities, suggesting a greater potential for collective cooperation to be formed in light traffic conditions. As traffic density increases, it becomes more difficult for vehicles to change lanes, so the potential for cooperation is reduced. 

Moreover, as observed in Figure~\ref{fig:surplus_headmap}, the surplus values consistently appear positive. According to Figure~\ref{fig:condition_cooperation}, a positive cooperation surplus may correspond to both 2-pipe and 1-pipe regimes, while collective cooperativeness is only attainable in 2-pipe regime. Consequently, in the following paragraphs, we examine the existence of collective cooperativeness by analyzing the empirical surplus values in more detail.

\noindent
\textbf{Collective cooperativeness}

Based on Steps~\ref{item:empirical_regime} and \ref{item:coop_existence} of Algorithm~\ref{algo:identification_procedure}, traffic regimes can be empirically differentiated by comparing the macroscopic speed data $\tilde{u}_1$ and $\tilde{u}_2$ obtained from Step~\ref{item:snapshot} with the computed 1-pipe speed $u^*$. Subsequently, the existence of collective cooperativeness is identified based on the proposed empirical existence condition, which involves two key factors: cooperation surplus and traffic regimes.

\begin{figure}[!ht] 
\centering
  \includegraphics[width=.8\linewidth]{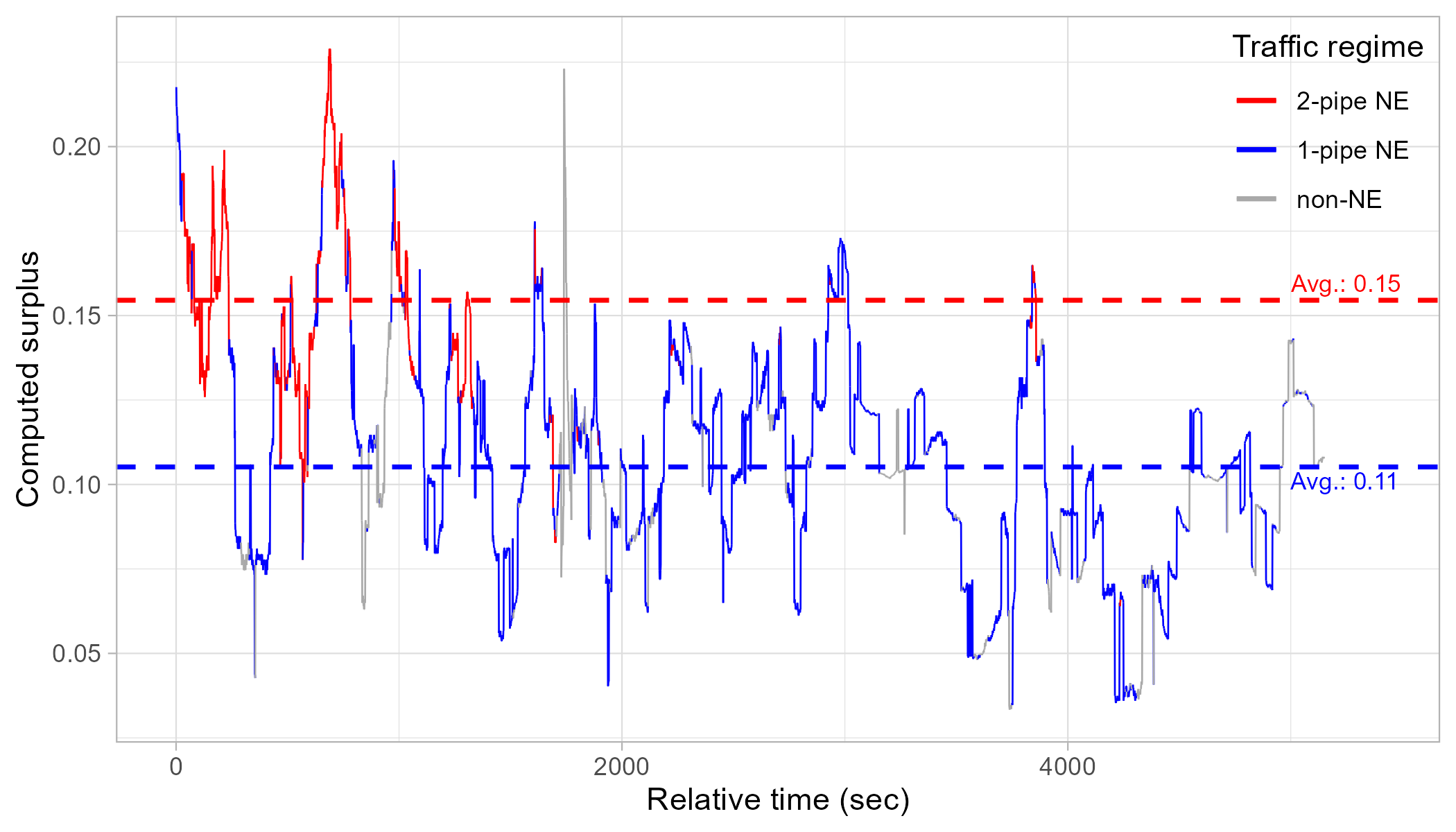}
  \caption{Cooperation surplus over time by traffic regimes} 
  \label{fig:surplus_time}
\end{figure}

We visualize the surplus values over time in Figure~\ref{fig:surplus_time}. The figure shows the type of Pareto-efficient NE (1-pipe or 2-pipe) attained when the surplus value changes. Specifically, the y-axis represents the value of surplus computed from the adapted game theoretical model, and the red/blue color shows which equilibria were observed. If neither of the conditions specified in (\ref{eq:regime_empirical}) are met, we call the state non-Pareto-efficient NE, labeled as the gray line. The arise of non-Pareto-efficient NE is because in real-world traffic, agents do not always reach or stay at an equilibrium state while they dynamically negotiate speeds.

Now we take a closer look at the two equilibria. Empirically, we find that the surplus value needs to be sufficiently large for the mixed traffic to reach 2-pipe Pareto-efficient NE. This can be seen from the average values of surplus in Figure~\ref{fig:surplus_time}, which is 0.15 for 2-pipe Pareto-efficient NE and 0.11 for 1-pipe Pareto-efficient NE. This offers an additional insight into our earlier theoretical model \citep{li2022equilibrium}. In the theoretical model, we have shown that a positive surplus value is a necessary condition for reaching 2-pipe Pareto-efficient NE, but it is not a sufficient condition. The present result is consistent with this theoretical characterization and further shows that empirically a larger surplus value more likely leads to 2-pipe Pareto-efficient NE. 

Quantitatively, we propose a metric to assess the probability of the emergence of collective cooperativeness in real-world traffic, which is written as,
\begin{equation}
P_{coop} = \frac{N_{2pipe}}{N_{tot}}
\end{equation}
where $P_{coop}$ represents the proportion of instances that are in the 2-pipe Pareto-efficient NE. It approximates the probability of attaining collective cooperativeness. The term $N_{2pipe}$ is the number of cases in the 2-pipe regime, and $N_{tot}$ is the total number of cases examined, including 2-pipe NE, 1-pipe NE, and non-NE. The resulting value of $P_{coop}$ is 13.83\%. This suggests that at the studied I-80 segment, even in states of semi-congestion or full congestion, there is a 13.83\% possibility that collective cooperativeness occurs.

Moreover, the upper horizontal line in Figure~\ref{fig:surplus_time} shows an average surplus value of 0.15. It implies that achieving a 2-pipe Pareto-efficient NE results in an average lateral space saving of 15\% compared to a non-Pareto efficient NE. Assuming the lateral space is fully utilized and the road capacity is 2200 vph/lane, this saving translates to an average of 330 vph/lane ($2200 \times 15\% = 330$) capacity improvement.

\begin{figure}[!ht]
    \centering
  \includegraphics[width=.5\linewidth]{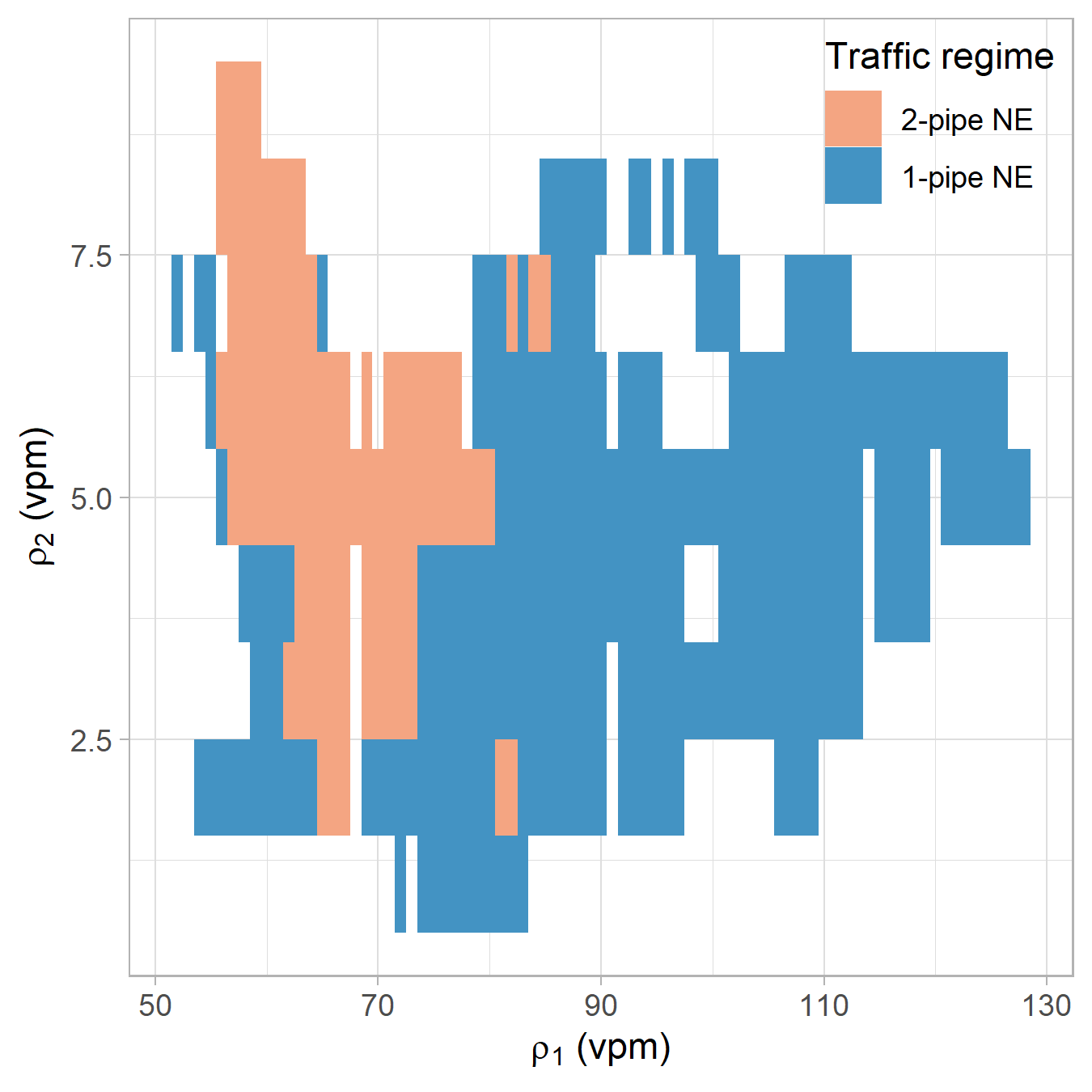}
  \caption{Traffic regime clusters}  
  \label{fig:heatmap_cluster}
\end{figure}

Figure~\ref{fig:heatmap_cluster} presents a visual depiction of the distribution of traffic regimes at different density levels under NE conditions. The 2-pipe regime is observed to occur more frequently at lower density levels, which aligns with higher surplus values in this regime. 

\begin{figure}[!ht]
    \centering
    \includegraphics[scale=0.55]{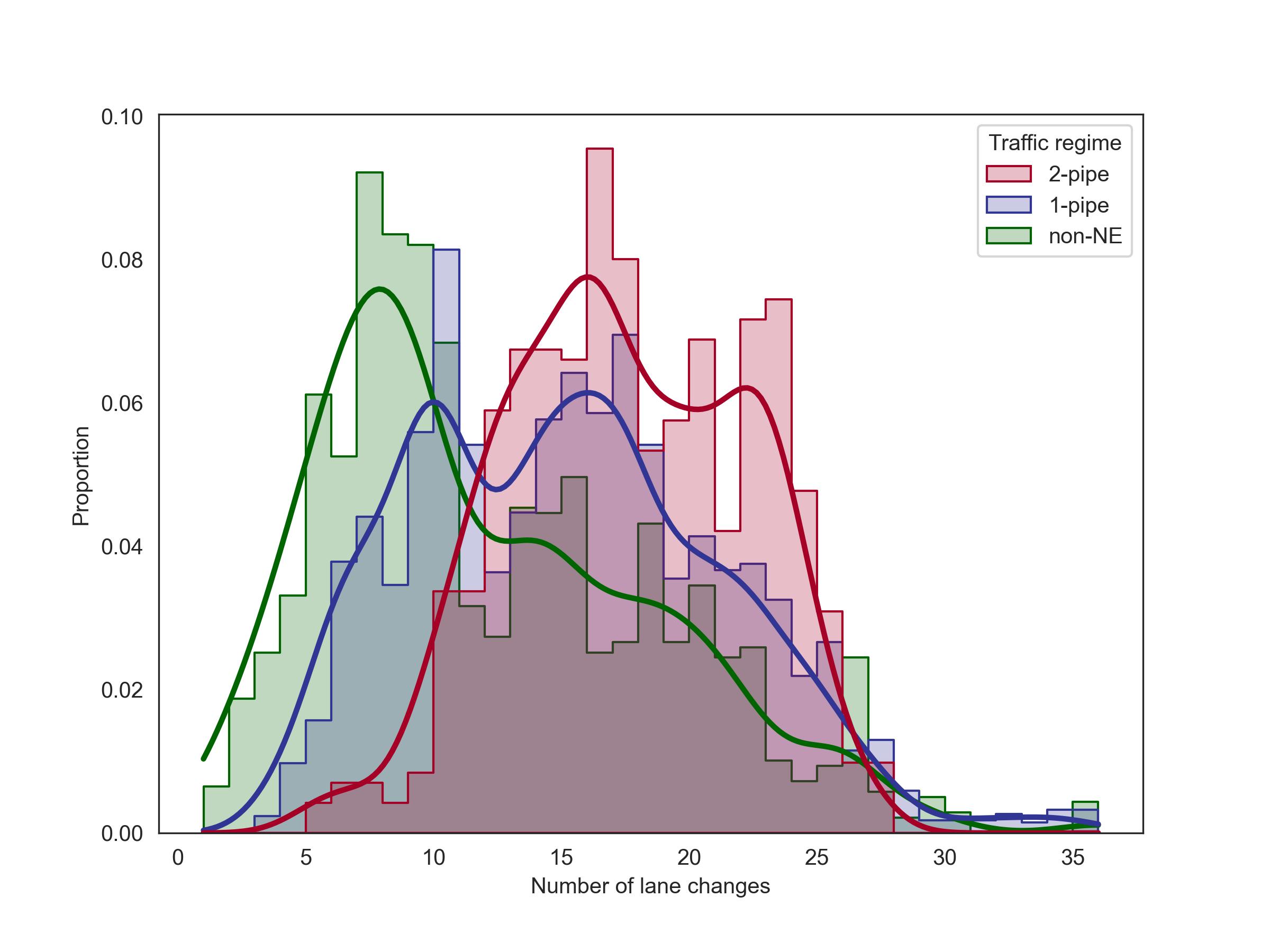}
    \caption{Lane change distribution by traffic regimes } 
    \label{fig:LC_freq}
\end{figure}

Moreover, we investigate the microscopic behaviors of agents to gain deeper insights into the factors influencing the disparities between different regimes. In Figure~\ref{fig:LC_freq}, we present the distribution of the number of lane changes across different traffic regimes. Lane changes are aggregated over one-minute intervals among all vehicles observed during that time period. For instance, if the figure shows a lane‐change count of 15 and there are 150 vehicles on the roadway, that translates to 0.1 lane changes per vehicle per minute. In other words, each vehicle typically changes lanes once every 10 minutes on average.
Importantly, the figure shows significant variations in lane change intensity with respect to regime changes. This observation is further supported by the Mann-Whitney U test~\citep{mann1947test}, which is employed to conduct pairwise comparisons of lane change distributions among the three regimes. The test statistics for all pairwise comparisons demonstrate values close to zero, indicating a statistically significant difference in the lane change intensities across regimes. 
Besides, the 2-pipe regime (i.e., the regime where collective cooperativeness exists) appears to have the largest lane change intensity while the non-NE has the smallest values. Our conjecture for this phenomenon is that a more active lane change behavior at the microscopic level can result in a more efficient utilization of lateral road spaces. Consequently, collective cooperativeness is more likely to emerge in this regime.

\subsubsection{Surplus split factor estimation result}\label{section:empirical_split_factor}

According to Figure~\ref{fig:identification_framework}, we have identified the scenarios where collective cooperativeness is present. Now we restrict our attention to these scenarios and characterize collective cooperativeness by estimating the surplus split factor $\lambda$. This is achieved by solving the inverse problem formulated in (\ref{eq:lambda_optm_loss}) and (\ref{eq:lambda_optm}). This is a one-dimensional and bounded problem, where $\lambda$ is the unknown parameter, constrained by $\lambda \in [0, 1]$. To solve this, we utilize Brent's method~\citep{brent1971algorithm}, which is effective for solving univariate minimization problems and locating the minimum of a function within a defined interval. This section corresponds to Step~\ref{item:identification} of Algorithm 1.

The dataset for the 2-pipe scenario is divided into a 70\% training set and a 30\% testing set. 
Regarding the weighting factors $w_1$ and $w_2$, two approaches have been considered: sample-based weighting, where the weights are proportional to the population size of each class, or class-based weighting, which treats each class as an equal player, regardless of its population size. The sample-based method risks biasing the estimation process toward the behavior of the more populous class, especially in cases of significant class size imbalance. This could misrepresent collective cooperativeness by overly focusing on a single class. Conversely, the class-based weighting eliminates population-based bias. As this aligns better with our objectives, we adopt the class-based approach, setting $w_1=w_2=0.5$ to ensure a balanced contribution from each class.
Furthermore, to mitigate the potential issue of over-fitting and selection bias during the estimation, we implement $k$-fold cross-validation \citep{stone1974cross}, a widely-used technique for model evaluation. This involves segmenting the training dataset into $k$ equally sized subsets, using $k-1$ of these subsets for model training and the remaining part for testing. This process is iterated $k$ times. For this study, we choose $k=10$.

Additionally, we adopt weighted MAEs to evaluate the estimation performance. The weighted MAE is computed as $MAE=\frac{1}{T}\sum_{t=1}^T \sum_{i=1}^2 w_i|\tilde{u}_{it}-\hat{u}_i(\rho_{1t},\rho_{2t};\hat{\lambda})|$, where $\tilde{u}_{it}$ is the $t$th macroscopic speed for class $i$ computed from (\ref{eq:snapshot_avg_speed}), $\hat{u}_i(\rho_{1t},\rho_{2t};\hat{\lambda})$ is the predicted speed for class $i$ based on the estimated surplus split factor $\hat{\lambda}$, and $T$ is the sample size (i.e., the total number of snapshots).

\begin{figure}[!ht]
\centering
\begin{subfigure}[t]{.45\textwidth}
  \centering
  \includegraphics[width=\linewidth, height=\textwidth]{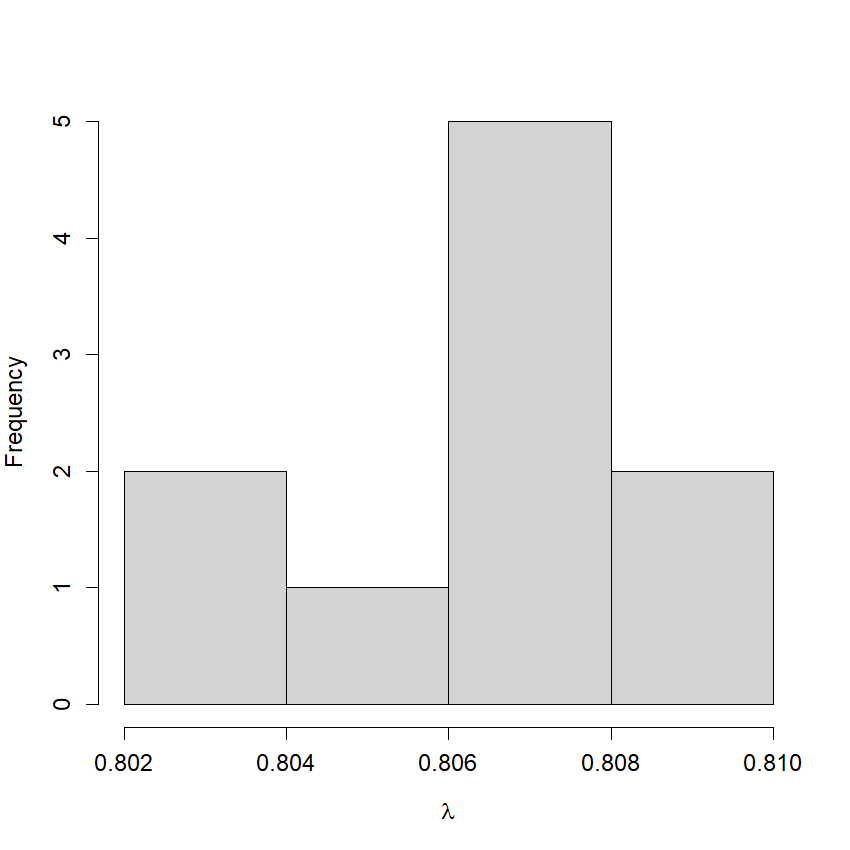}
  \caption{Distribution of the estimated surplus split factors}
  \label{fig:lambda_dist}
\end{subfigure}
\begin{subfigure}[t]{.45\textwidth}
  \centering
  \includegraphics[width=\linewidth, height=\textwidth]{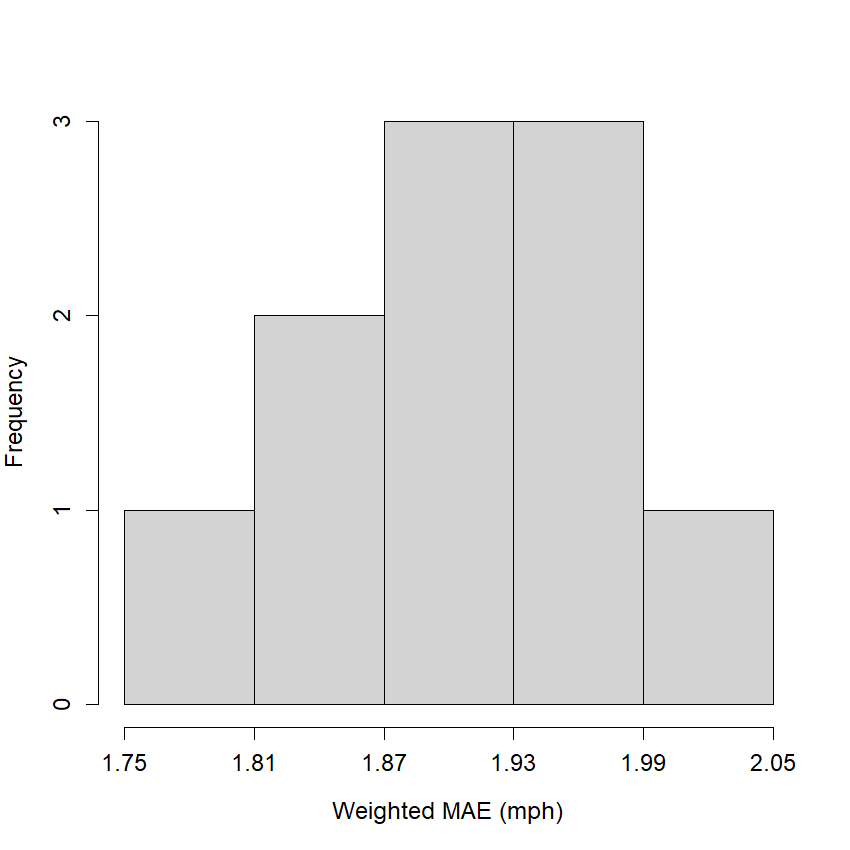}
  \caption{Distribution of the weighted MAEs}
  \label{fig:lambda_mae}
\end{subfigure}
\vspace{0.2cm}
\caption{Distribution of estimated surplus split factors and weighted MAEs in 10-fold cross validation}
\label{fig:lambda_obj_validation}
\end{figure}

Figure~\ref{fig:lambda_obj_validation} shows the estimated values of the surplus split factors and the corresponding weighted MAEs obtained through 10-fold cross-validation. The values of $\lambda$ and weighted MAEs across each iteration demonstrate both closeness and consistency, ruling out outcome-of-chance and underscoring the reliability of the split factor estimation process. Additionally, this uniformity suggests the model's transferability to other datasets and real-world scenarios.
The final result for the estimation is shown in Table~\ref{table:lambda}. The table outlines the class-specific MAEs and the weighted MAE, all of which are relatively low. The low errors indicate that the macroscopic speed function, as defined in (\ref{eq:payoff}), and the estimated split factor, are consistent with real-world data. 

\begin{table}[h]
    \centering
    \caption{Surplus split factor estimation result}
    \begin{tabular}{cccc}
    \toprule
     $\hat{\lambda}$ & Class 1 MAE (mph) & Class 2 MAE (mph) & Weighted MAE (mph) \\ 
     \midrule
     0.8067	& 2.32 & 1.45 & 1.89  \\ 
     \bottomrule
    \end{tabular}
    \label{table:lambda}
\end{table}

\begin{figure}[!ht]
    \centering
    \includegraphics[scale=0.4]{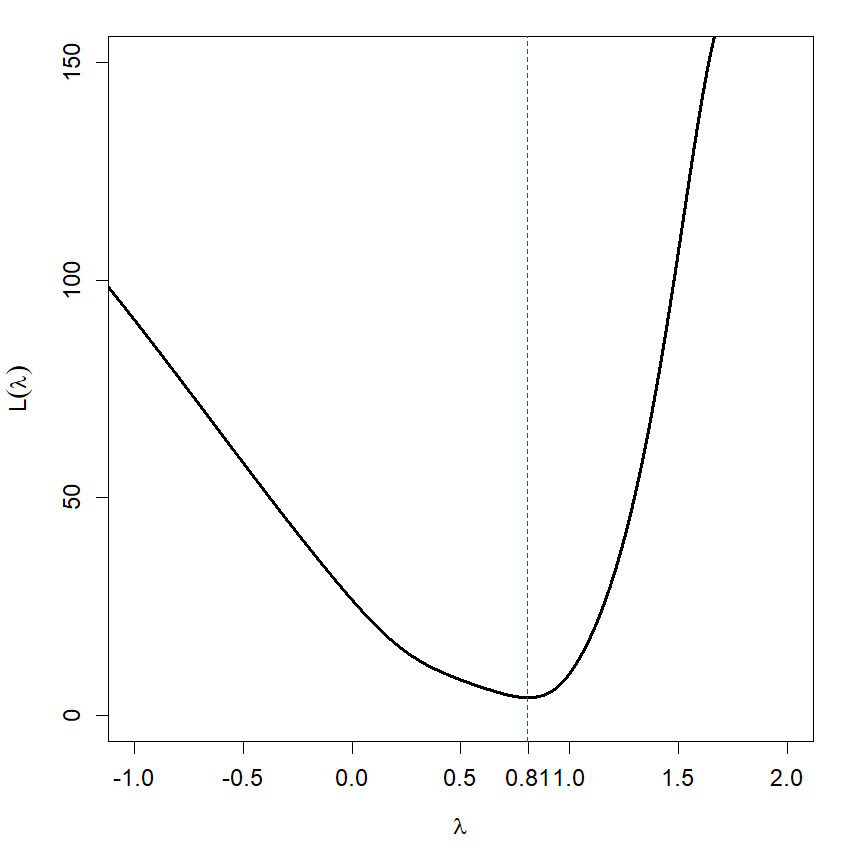}
    \caption{Value of loss function and identified $\hat{\lambda}$}
    \label{fig:lambda_obj}
\end{figure}

Figure~\ref{fig:lambda_obj} depicts the values of the objective function for various surplus split factors, with the optimal solution highlighted in red. The solution lies within the range of [0, 1] with optimal value $\hat{\lambda}=0.8067$, signifying mutual benefits from cooperation for both vehicle classes. Moreover, the figure shows that the objective function's slope to the right of the optimal solution is steeper than on the left. This implies that the shift of equilibrium towards a smaller split factor is more likely. We interpret this as a higher tolerant level among the population of cars to receive less cooperation surplus, possibly due to that for each car, the reduction of normalized surplus is relatively small. 

We further compute the normalized surplus split factor by (\ref{eq:lambda_optm_normalize}). For passenger cars, the PCE is 1, and for trucks, the PCE is set to 1.5, reflecting the congested freeway scenario in this study~\citep{FHWA2000}. The filtered dataset includes 1401 cars and 39 trucks, resulting in normalized factors of $\tilde{\lambda}_1 = 0.84$ and $\tilde{\lambda}_2= 4.82$. Besides, the computed equity metric from (\ref{eq:equity_metric}) is $\zeta = 3.98$. These results indicate that the allocation of cooperation surplus is inequitable, with trucks gaining a larger share of the cooperation surplus and thus obtaining more benefits from collective cooperation. This could be attributed to the larger size and weight of trucks, which allows them to occupy road space more aggressively. The larger headway of trucks may also contribute to the inequity.

In summary, in this section, we apply our framework to a real-world dataset to investigate the existence of collective cooperativeness and characterize it in mixed human-driven traffic. Our analysis yields several insights and verifies the effectiveness of our approach.  
Firstly, in Section~\ref{section:empirical_nominal}, we identify nominal speed functions. We also demonstrate that the scaling parameters provide a reasonable approximation to real-world data. This allows us to adopt one nominal speed function for each vehicle class, while accounting for the type-dependency with the leading vehicle class.
Secondly, in Section~\ref{section:empirical_surplus}, we confirm the existence of collective cooperativeness. We also present the relationship between traffic regimes and cooperation surplus from different perspectives, and measure the likelihood that collective cooperativeness exists.
Lastly, in Section~\ref{section:empirical_split_factor}, we identify the surplus split factor between the two classes of vehicles, which implies how the benefits of cooperation is allocated between vehicle classes.

\section{Conclusion}\label{section:conclusion}

How traffic agents interact, e.g., how they compete for lateral spaces, plays a key role in shaping the properties of traffic flow. This is especially true when heterogeneous traffic agents are considered, such as in mixed autonomy traffic flow. While the understanding of traffic flow from the dynamic system angle is relatively thorough, the understanding of it from the multi-agent behavioral perspective is relatively nascent. Among others, detection and modeling the emergence of cooperation in traffic flow is still an open problem. This paper contributes to the understanding of collective cooperation in traffic flow from an empirical perspective.

The present work is the first one that empirically identifies and characterizes collective cooperativeness in mixed traffic. One major technical contribution of this work is a new framework to define and identify the collective cooperativeness of heterogeneous traffic agents from real-world trajectory data. This framework complements our theoretical model~\citep{li2022equilibrium} to enable inferences from vehicle trajectory data by leveraging information at two scales, i.e., the microscopic and macroscopic scales. By doing so, we can identify parameters that characterize collective cooperation.

We apply the proposed framework to the NGSIM I-80 trajectory data. After careful data processing, our major findings are threefold. Firstly, the scaling property holds for the classes of cars and trucks. Secondly, the cooperation surplus exists in the human-driven traffic, and a larger surplus value tends to more likely lead to Pareto-efficient NE. Thirdly, the surplus split factor is 0.81, and the loss function is asymmetric with respect to this value, indicating varying sensitivity levels between cars and trucks to the changes in road share allocation. Besides, the equity metric is 3.98, indicating an inequitable allocation on the benefits of cooperation, with trucks securing more benefits than cars.

We envision several relaxations and enhancements in future works. Firstly, this framework is based on the assumption that all agents are endowed with equilibrium speed functions. This assumption, while simplifying the model and allowing for better analytical tractability, is an idealized representation of real-world traffic. It is imperative to define and identify collective cooperation in more general settings, taking detailed agent level dynamics into consideration. Secondly, this framework assumes each class of agents follows the same behavior rules. Even though this is a stylized assumption in multi-class traffic modeling literature, in reality, how to group traffic into different classes is not a straightforward problem. 
Thirdly, it is also interesting to consider the cooperation concept in mixed traffic with more than two classes of traffic agents. Such an extension would involve developing multi-class analytical models to handle the increased complexity of interactions among multiple classes and adapting the identification framework accordingly.
Last but not least, it is worthwhile to explore how collective cooperation can inform the control of mixed autonomy traffic and the design of AV behaviors to benefit all road users.

\section*{Acknowledgment}{%
This research is supported by NSF grant ``EAGER: Collaborative Research: Fostering Collective Rationality Among Self-Interested Agents to Improve Design and Efficiency of Mixed Autonomy Networks and Infrastructure Systems'' (Award numbers: 2437983, 2437982), and CTech grant ``Design Automated Vehicle Behaviors in Mixed Traffic Flow".
}%

\bibliographystyle{plainnat} 
\bibliography{references}  






\end{document}